\documentclass[oneside]{amsart}

\usepackage{geometry}
\usepackage{float}
\usepackage{graphicx}
\usepackage{amssymb}

\theoremstyle{plain}
\newtheorem{thm}{Theorem}[section]
  \theoremstyle{definition}
  \newtheorem{defn}[thm]{Definition}
  \theoremstyle{plain}
  \newtheorem{algorithm}[thm]{Algorithm}
  \theoremstyle{remark}
  \newtheorem*{acknowledgement*}{Acknowledgement}

\newcommand{\mean}[1]{\left\langle #1 \right\rangle}
\newcommand{\ra}{\rightarrow}

\makeatletter
\newcommand{\url}[1]{%
  \texttt{\language\@M\hyphenchar\the\font=`\/ #1}}%
\makeatother

\begin{document}

\title{Dual Computations of Non-abelian Yang-Mills on the Lattice}
\author{J. Wade Cherrington$^1$}
\author{J. Daniel Christensen$^2$}
\author{Igor Khavkine$^3$}

\maketitle

\vspace*{-10pt}
\begin{center}
\small
$^{1,3}$Department of Applied Mathematics, University of Western 
Ontario, London, Ontario, Canada\\
$^2$Department of Mathematics, University of Western Ontario, London, Ontario, Canada\\
E-mail: jcherrin@uwo.ca, jdc@uwo.ca, and ikhavkin@uwo.ca
\end{center}

\begin{abstract}
In the past several decades there have been a number of proposals
for computing with dual forms of non-abelian Yang-Mills theories on
the lattice. Motivated by the gauge-invariant, geometric picture offered
by dual models and successful applications of duality in the $U(1)$
case, we revisit the question of whether it is practical to perform
numerical computation using non-abelian dual models. Specifically,
we consider three-dimensional $SU(2)$ pure Yang-Mills as an accessible
yet non-trivial case in which the gauge group is non-abelian. Using
methods developed recently in the context of spin foam quantum gravity,
we derive an algorithm for efficiently computing the dual amplitude
and describe Metropolis moves for sampling the dual ensemble.
We relate our algorithms
to prior work in non-abelian dual computations of Hari Dass and his collaborators,
addressing several problems that have 
been left open. We report results of spin expectation value computations
over a range of lattice sizes and couplings that are in agreement
with our conventional lattice computations. We conclude with an outlook
on further development of dual methods and their application to problems
of current interest.
\end{abstract}

\section{Introduction}

In this work we describe recent progress in lattice gauge theory (LGT)
computations in a model dual to pure Yang-Mills theory. The dual model
we consider is based upon the character expansion of the amplitude
at each fundamental plaquette, a procedure that has been known since
the early days of lattice gauge theory~\cite{Kogut83,Osterwalder,Savit}.
To date, the most common use of this duality transformation has been
in the strong-coupling expansion, as described for example in~\cite{ItzyksonDrouffe,KogutStephanov,MontvayMunster}
and references therein. Because of this historical association with 
strong-coupling approximations, we should emphasize here that it 
is the exact dual model that we compute with, sampling from the 
full space of dual configurations to find the expectation values of observables. The first steps towards
such dual computations in the non-abelian case were taken by Hari Dass
et al.\ in~\cite{Dass83,Dass94,DassShin}; more recently, interest
in dual computations of lattice Yang-Mills theory has grown within
the spin foam community~\cite{Conrady,OecklPfeiffer,PfeifferDual}. 

\begin{figure}[H]
\includegraphics[scale=0.31]{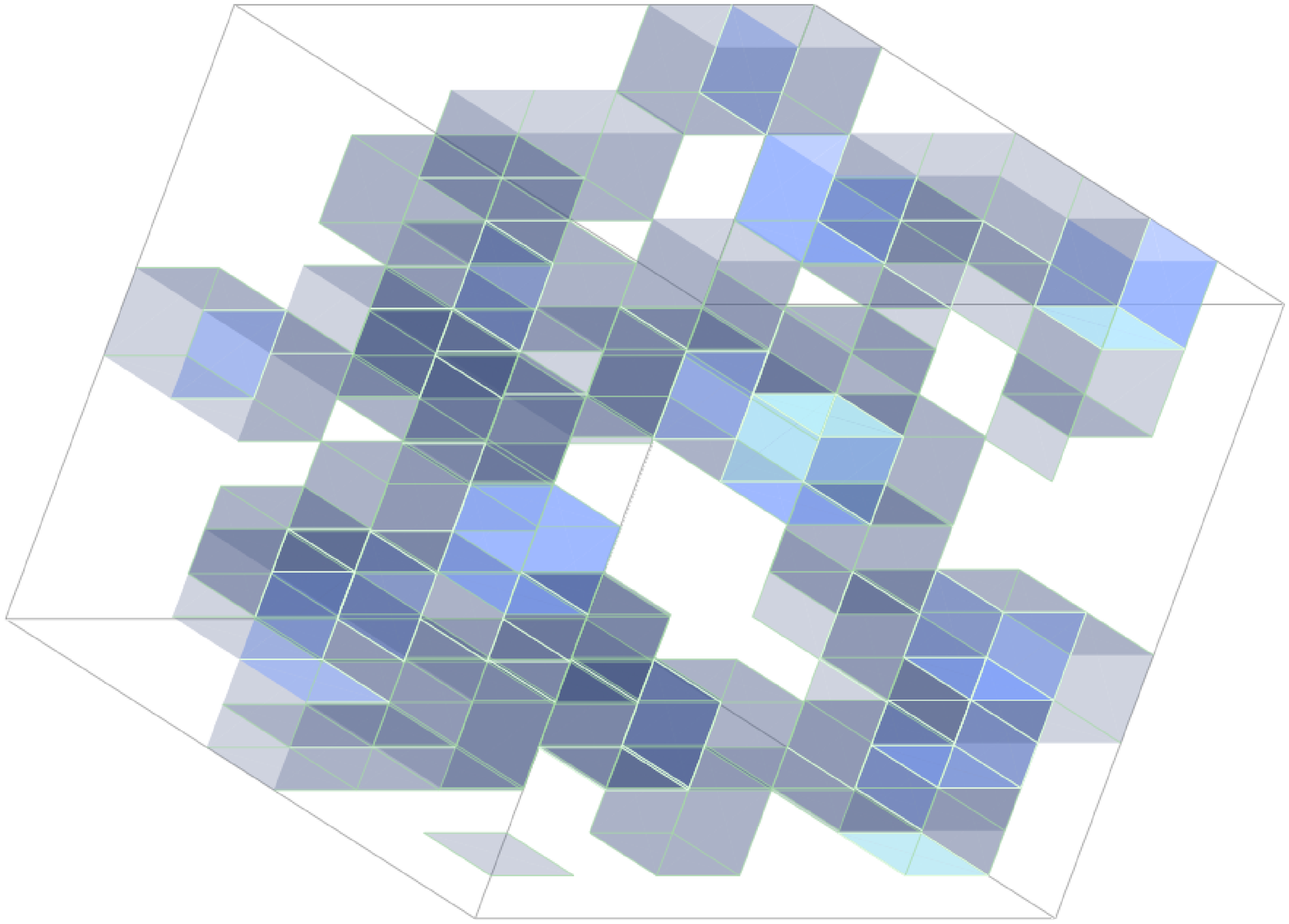}~~~\includegraphics[scale=0.28]{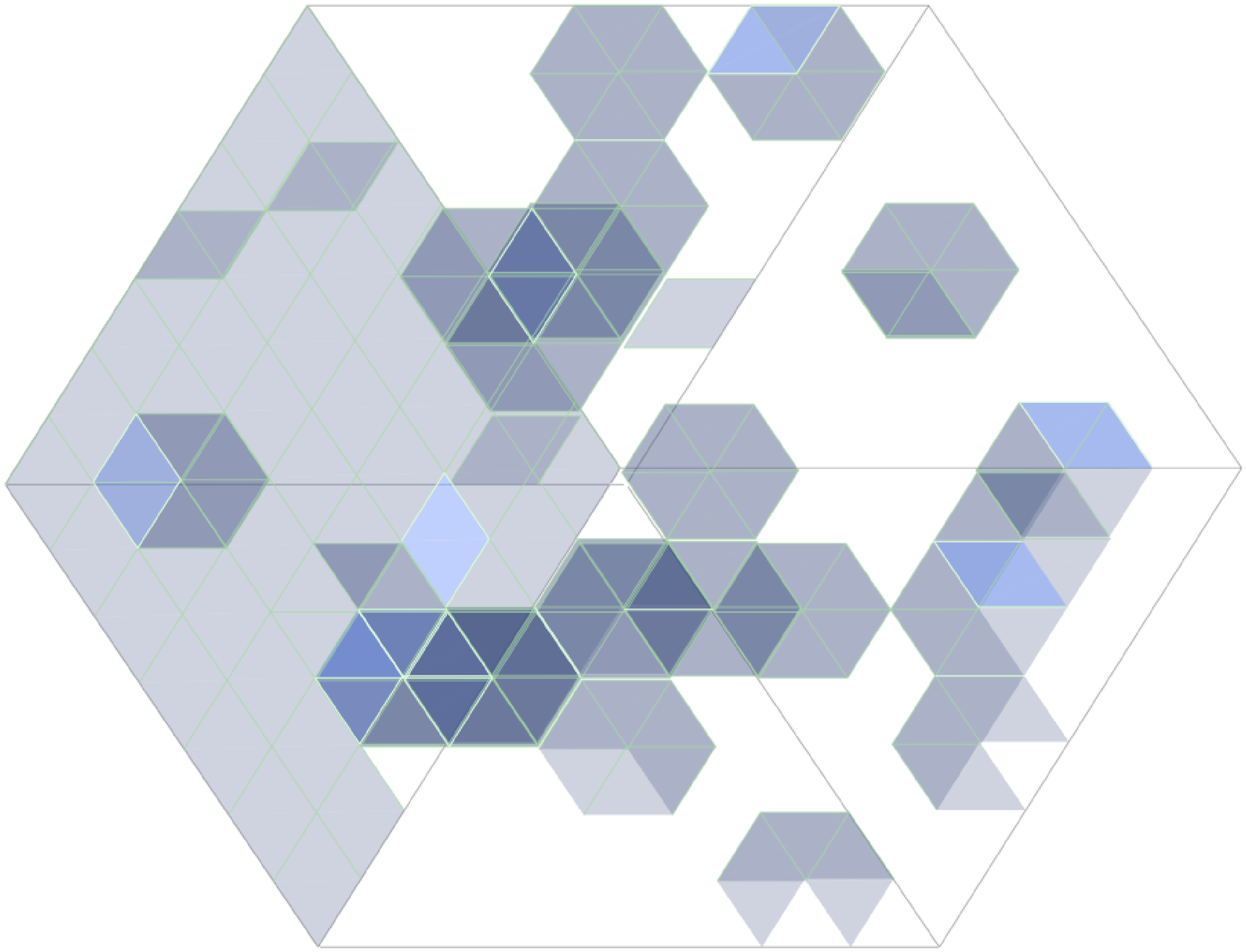}

\caption{Two examples of dual configurations on a $6^{3}$ lattice. Opposite
sides of the lattice are identified. In the right-hand figure a sheet
of flux spanning the lattice is present, allowed due to the non-trivial homology of the
3-torus. In the left-hand figure there are no such sheets present.}
\label{fig:visuals}
\end{figure}

Perhaps the most significant change in passing to the dual model is
that the variables include discrete labels assigned
to lattice plaquettes. This aspect alone gives a very different character
to the computation when compared to the conventional formulation of
LGT, where continuous group-valued variables are assigned to lattice
edges. Another significant difference is the fact that the discrete 
labels must satisfy certain constraints,
making the choice of Metropolis moves non-trivial (see Section~\ref{se:review}).
The allowed configurations can be viewed
as closed, branched surfaces colored by the irreducible representations
of the gauge group. To illustrate, some renderings of non-zero dual
configurations are shown in Figure~\ref{fig:visuals}. 

As general motivation for the dual approach, we observe that the dual
configurations have a gauge-independent meaning rooted in the statistical
geometry of the two-dimensional branched surfaces, providing a compelling
geometric picture for the evolution of the physical degrees of freedom.
For example, in the strong-coupling limit, the dual picture provides
a straightforward analytic proof of confinement (see, e.g.,~\cite{MontvayMunster}
and references therein). By making topological excitations manifest,
a dual model is well suited to evaluating proposed mechanisms of quark
confinement, such as the dual superconductor picture. This advantage
has already been demonstrated in the abelian $U(1)$ case~\cite{Jersak,Panero2004,Panero2005,PollyWiese,Zach96,Zach98,Zack98PhysRev}
and has also been remarked upon in the non-abelian case~\cite{OecklPfeiffer}.
Recently, effective theories have been derived directly from the dual
theory~\cite{ConradyGluons}, and so we believe a computational framework
for numerical computations within the dual theory is timely. 

While results have been reported with dual models in the $U(1)$ 
theory, the non-abelian case has presented a greater computational challenge.
As a testing ground for dual non-abelian simulation algorithms, we
focus in the present work on $SU(2)$ pure Yang-Mills in three space-time
dimensions.  The observable we study is the average spin (see Section~3)
which is convenient for testing as it takes a simple form on both sides
of the duality.
In describing our algorithm and results, we address potential
problems for non-abelian dual computations that were discussed in
the work of Hari Dass et al.~\cite{Dass94,DassShin}, the most critical
of which were the construction of ergodic moves and a sign problem.
The results obtained with our Metropolis algorithm appear to overcome
these difficulties for the range of lattice sizes and coupling
constants presented here. With the current algorithm, a difficulty (discussed in 
Section~4.2) emerges at weak coupling, however our expectation is that it
will be resolved with a more refined form of the algorithm. 

Although there is an intriguing possibility that dual algorithms may
eventually outperform their conventional equivalents in certain contexts,
the present work focuses on the general features of computing with
dual algorithms and verifying their correctness against a conventional
lattice code. In future work, we will address more optimized implementations
of the algorithms described here, with the ultimate goal of providing
results in four dimensions with lattice sizes typical of contemporary
lattice QCD.
An important next step will be to extend our methods to Wilson loop
observables crucial to the study of quark confinement and glueball
spectra; this work is currently in progress~\cite{CherringtonWilson}.

The paper is organized as follows. In Section 2, we define the dual
model under study and describe a Metropolis algorithm for numerical
computations. In Section 3, we describe computations on a $2^{3}$
lattice that allow us to verify threefold agreement (within statistical
error) between dual Metropolis, conventional Metropolis and the exact
partition function. In Section 4, we describe our computations for lattice
sizes of $4^{3}$, $8^{3}$, and $16^{3}$ and report agreement between
dual and conventional Metropolis results, again within statistical error.
Section 5 provides a brief outlook on generalizing the algorithm and
Section 6 presents our conclusions. Some details relevant for performing
the computations are given in the Appendix, which also shows how the
dual amplitude formula used by Hari Dass~\cite{Dass94,DassShin},
originally given in~\cite{Anishetty93} and discussed in~\cite{DiaPet99,Halliday95},
fits into the spin foam formalism used to develop our algorithm.

\section{Review of The Dual Model And Algorithm Construction}
\label{se:review}

In this section we define the dual model for $SU(2)$ pure Yang-Mills
on a hyper-cubic lattice and describe a Metropolis algorithm for performing
dual computations. Of particular relevance to our present work is the application of diagrammatic
methods and spin foam formalism, as described for example in~\cite{Conrady,OecklPfeiffer,Oeckl},
a critical feature being the expression of group integrals of representation
matrices and their contractions as sums over intertwiners%
\footnote{An intertwiner is a map between representations of a group that commutes
with the action of the group.%
}.  While the spin foam formalism for dual models
is a more recent development, the construction of exact dual non-abelian models has a longer 
history~\cite{Anishetty90, Anishetty91,Anishetty93,Halliday95,DassNIKHEF,DassNuclB,DassPhysB}.
Traditional derivations of the duality transformation
and their use in strong-coupling expansions can be found in the texts
\cite{ItzyksonDrouffe,MontvayMunster}.

\subsection{Review of pure Yang-Mills theory on the lattice}
\label{sse:review}

First we recall  the Euclidean partition function of pure Yang-Mills
theory in $D$ dimensions, with gauge group $G=SU(N)$, where $N\geq2$ (we
shall later specialize to the $SU(2)$ case).  It takes the form
\begin{equation}\label{eq:YMpartfunc}
	\mathcal{Z} = \int \mathcal{D}A\, \exp(-S),
\end{equation}
with $A^a_\mu$ the gauge field, $S$ the action functional, and
$\mathcal{D}A$ the functional integration measure. In the continuum
version of the theory the standard action functional is
\begin{equation}
	S \equiv S[A] = \frac{1}{4g^2} \int d^Dx\, F^a_{\mu\nu} F_a^{\mu\nu},
\end{equation}
where $F^a_{\mu\nu}$ is the field strength tensor and $g$ the continuum 
coupling. Unfortunately, the continuum functional measure
$\mathcal{D}A$ is not well-defined.

One way to give the above path integral rigorous meaning and, at the
same time, make it amenable to computational treatment, is to put the
theory on a discrete finite lattice. The simplest variant uses a
hyper-cubic lattice. Let $E$ and $P$ denote respectively the sets of
edges and plaquettes of a hyper-cubic lattice in $D$ dimensions.  The
gauge field $A$ is replaced by gauge group elements $g_e$ assigned to
each oriented lattice edge $e\in E$. The same
edge with opposite orientation gets $g_e^{-1}$ instead of $g_e$. The functional integral measure can now
be replaced by an integral over the
product of $|E|$ copies of $G$ using Haar measure:
\begin{equation}
	\mathcal{D}A \equiv \prod_{e\in E} dg_e.
\end{equation}
At the same time, the action functional is replaced by a discretized
version, $S\equiv S[g]$, that must reproduce the continuum action $S[A]$
as the lattice spacing is taken to zero. The discretized action is
usually split into a sum over plaquettes, $S[g] = \sum_{p\in P} S(g_p)$,
where the group element $g_p$ is the holonomy around an oriented 
plaquette $p$.  That is, $g_p = g_1 g_2 g_3 g_4$, where $g_i$ is
either the group element assigned to the $i$th edge of $p$ or its
inverse if the orientations of $p$ and the $i$th edge are opposing.
This yields the conventional lattice partition function
\begin{equation}\label{eq:conventional}
	\mathcal{Z} = \int \prod_{e\in E} dg_e\, e^{-\sum_{p\in P} S(g_p)}.
\end{equation}

There are many candidate discretized plaquette actions $S(g_p)$.
While the Wilson action~\cite{Wilson} is perhaps the most well-known in
conventional LGT (it was also used in the dual computations of~\cite{Dass83,Dass94,DassShin}),
 a variety of actions $S(g_{p})$ leading to the correct continuum limit are
known and have been used in the literature~\cite{Horn,Lang1,Lang2}.
In the present work, we use the \emph{heat kernel action}~\cite{Menotti};
in the dual model this action leads to plaquette factors that are
particularly easy to compute. The heat kernel action (at lattice coupling
$\gamma$) for a fundamental plaquette $p$ and plaquette holonomy
$g_{p}$ is \begin{equation}
e^{-S(g_{p})}=\frac{K(g_{p},\frac{\gamma}{2}^{2})}
{K(I,\frac{\gamma^{2}}{2})},
\label{eq:heatkernel}
\end{equation}
where the heat kernel $K$, which is a function of a group element $g$ and of
a ``time'' parameter $t$, satisfies a diffusion type differential
equation 
\begin{equation}\label{eq:K}
\frac{\partial}{\partial t}K(g,t)=\Delta K(g,t),\quad K(g,0)=\delta_{I}(g) .
\end{equation}
Here $\Delta$ is the Laplace-Beltrami operator on $G$ and $\delta_{I}$
is the delta function at the group identity $I$. The denominator in
(\ref{eq:heatkernel}) represents a normalization of the partition
function in which flat holonomies ($g_{p}=I$) are assigned an amplitude
of unity. We shall follow the common practice of discussing
the phase structure of a lattice theory using the $\beta$ parameter
$\beta=\frac{4}{\gamma^{2}}$.

We now turn to the definition of the dual model for the specific case
of $G=SU(2)$ pure Yang-Mills in three dimensions. Starting from the
conventional formulation of the lattice partition function $\mathcal{Z}$
given in~(\ref{eq:conventional}) above, the duality transformation
can be applied (see Appendix~\ref{sse:derivation}) to
yield the following expression for $\mathcal{Z}$ in terms of
the dual variables:
\begin{equation}
\mathcal{Z}=\sum_{j}\left(\sum_{i}
	\prod_{v\in V}{18j}^v(i_{v},j_{v})
	\prod_{e\in E} N^e(i_e,j_e)^{-1}\right)
\left(\prod_{p\in P}e^{-\frac{2}{\beta}j_{p}(j_{p}+1)}(2j_{p}+1)\right).
\label{eq:statesum}
\end{equation}
Here $V$ denotes the vertex set of the lattice, while the summations
over $i$ and $j$ range over all possible edge and plaquette labellings,
respectively. A plaquette labelling $j$ assigns an irreducible
representation of $SU(2)$ to each element of $P$. These representations
are labelled by non-negative half-integers (we will denote this set 
by $\frac{1}{2}\mathbb{N}$) and are referred to as \emph{spins};
a labelling $j$ is thus a map $j\colon P\to\frac{1}{2}\mathbb{N}$. 
An edge labelling $i$, on the other hand, is valued in a basis of maps that intertwine the 
representations of the plaquettes incident on the same edge.  In our present case, the choice of basis corresponds to a grouping of the four incident plaquette spins into two pairs.
When such an edge splitting has been made, the intertwiners may also be labelled by
spins, as described just before Definition~\ref{defn:admissible-spin-foam} and in 
Appendix~\ref{sse:algs}.
Different choices of splitting can be made, but, some are more 
computationally efficient than others. In writing~(\ref{eq:statesum}), we assume
a fixed choice of splitting has been made and so an edge
labelling is a map $i\colon E\to\frac{1}{2}\mathbb{N}$. 

In the first pair of parentheses of~(\ref{eq:statesum}), there is
a product of $18j$ \emph{symbols}, each of which is a function
of the 18 spins which label the $12$ plaquettes and $6$ edges
incident to a vertex $v$;  we denote the spins which appear
by $j_v$ and $i_v$.
Next to it is a product of edge normalizations $N^e$ depending on the
edge spin $i_e$ and on the four spins $j_e$ labelling the plaquettes
incident on $e$. 
It is important to recognize that the $18j$ symbol and the
normalization factors $N^e$ are purely 
representation-theoretic quantities (independent of the action chosen) and that, 
from a computational viewpoint, they represent the non-trivial part of the amplitude 
evaluation.
Efficient algorithms can be found (using diagrammatic techniques similar
to those used in~\cite{ChristensenEgan}) for computing the $18j$ symbols and edge normalizations.
Two of these are reviewed in Appendix~\ref{sse:algs}.

In the second parentheses of~(\ref{eq:statesum}) there is
a product of factors depending on $j_{p}$ only; these arise from
the character expansion coefficients of the heat kernel action~(\ref{eq:heatkernel})
and are clearly straightforward to compute. 

For the purposes of this paper, we define a \emph{spin foam} to be
an assignment of spins and intertwiners to the plaquettes and edges
of the lattice%
\footnote{In spin foam quantum gravity, the 2-cells and 1-cells of a simplicial complex
are typically used, rather than the plaquettes and edges of a cubic
lattice. The general definition of spin foam was introduced by Baez
in~\cite{BaezSpinFoam}.%
}, respectively. 
We define \emph{supported spin foams} to be those with non-vanishing
amplitude, and denote the set of them by $\mathcal{F}^{+}$. In terms of the
supported spin foams we can write~(\ref{eq:statesum}) 
as:
\begin{equation}
\mathcal{Z}=\sum_{f\in\mathcal{F^{+}}}\mathcal{A}(f)=
\sum_{f\in\mathcal{F^{+}}}\left(\prod_{v\in V}18j^v(i_v,j_v)
\prod_{e\in E} N^e(i_e,j_e)^{-1}\right)
\left(\prod_{p\in P}e^{-\frac{2}{\beta}j_{p}(j_{p}+1)}(2j_{p}+1)\right).
\end{equation}
It turns out that the spin foams that actually make
a non-zero contribution to $\mathcal{Z}$ are highly constrained,
a fact that is not manifest in~(\ref{eq:statesum}) where every possible
edge and plaquette coloring contributes a term. Moreover, once
the plaquette spins have been specified, there is a limited range
of intertwiner spins that give a non-zero amplitude. The details of
these constraints will be discussed in Section 2.2. 

The essential challenge for performing Metropolis simulations
with this model is to find a set of moves that connect all
the supported spin foams $\mathcal{F^{+}}$. To see why this is
a non-trivial task, observe that if one has any supported spin
foam and adds or removes a half-unit of spin from a single plaquette,
the result will be a spin foam with vanishing amplitude; this follows 
immediately from the parity condition (see Definition
2.1 below), enforced by the $18j$ symbol. Changing single plaquettes by 
two half-units of spin preserves parity but restricts one to a single parity class. 
Thus a move involving a single plaquette is ``too local''
and so moves involving multiple plaquettes and edges simultaneously are needed.  
We now turn to this problem.

\subsection{Algorithm Construction --- Ergodic Moves}

We first define the set of \emph{admissible plaquette configurations.}
These configurations are precisely those plaquette labellings that satisfy \emph{edge admissibility}
at every edge of the lattice:

\begin{defn}[Edge admissibility] 
The spins assigned to plaquettes incident
to an edge are said to be \emph{edge admissible} if the \emph{parity}
and \emph{triangle inequality} conditions are satisfied. 
Writing $j_1$, $j_2$, $j_3$ and $j_4$ for the four spins incident to a
given edge, these conditions are:
\begin{enumerate}
\item \textbf{Parity}: \[
j_{1}+j_{2}+j_{3}+j_{4}\text{ is an integer.}\]

\item \textbf{Triangle Inequality}: for each permutation $\left(k,l,
  m, n\right)$ of $\{1, 2, 3, 4\}$ we have
\[
  j_{k} + j_{l} + j_{m} \geq j_{n} .
\]
\end{enumerate}
These conditions are equivalent to the existence of a non-zero
invariant vector in the $SU(2)$ representation 
$j_1\otimes j_2 \otimes j_3\otimes j_4$.
\end{defn}

If a splitting has been made for the edge, say with
$j_1$ and $j_2$ on one side and $j_3$ and $j_4$ on the
other, then the triangle inequality can be written in the
less symmetric form
\begin{equation}
\left|j_{1}-j_{2}\right|\leq j_{3}+j_{4},\,\,\left|j_{3}-j_{4}\right|\leq j_{1}+j_{2}.
\end{equation}
Representation-theoretically, the less symmetric form says
that there exists a non-trivial intertwiner between 
$j_1\otimes j_2$ and $j_3\otimes j_4$.
A basis of such intertwiners is labelled by spins which
match parity with $j_{1}+j_{2}$ and
$j_{3}+j_{4}$ and fall into the specific range defined below. This leads
us to the following definition of an \emph{admissible spin foam}:

\begin{defn}[Admissible spin foam]\label{defn:admissible-spin-foam}
A spin foam is \emph{admissible} if
and only if:
\begin{enumerate}
\item The assignment of spins to plaquettes is everywhere edge admissible.
\item For every edge $e\in E$,  $i_{e}$ satisfies
\[
i_{e}+j_{1}+j_{2} \text{ and } i_{e}+j_{3}+j_{4} \text{ are integers} 
\]
and
\[
i_{e}\in I_{e}\equiv[\left|j_{1}-j_{2}\right|,j_{1}+j_{2}]\cap[\left|j_{3}-j_{4}\right|,j_{3}+j_{4}],
\]
where $\left[a,b\right]\equiv\left\{ j\in\frac{1}{2}\mathbb{N}\mid a\leq j\leq b\right\} $.
\end{enumerate}
We call the interval $I_e$ the \emph{range} of the triangle inequalities
at edge $e$. It is guaranteed to be non-empty by edge admissibility.
We denote the set of admissible spin foams by $\mathcal{F}^{A}$.
\end{defn}

At this point we observe that while any supported spin foam is admissible,
the converse does not necessarily hold, as the amplitude can vanish
despite admissibility conditions being satisfied. We define \emph{exceptional
spin foams} to be admissible spin foams of vanishing amplitude and
denote them by $\mathcal{F}^{E}\equiv\mathcal{F}^{A}\setminus\mathcal{F}^{+}.$
In practice, exceptional spin foams are rarely encountered during simulation.
If one assumes that exceptional spin foams are sufficiently isolated
(i.e.\ do not form surfaces that separate the configuration space)
then the question of ergodicity will not be affected, i.e.\ it will
be sufficient to show ergodicity of moves on admissible spin foams.
In what follows, we shall make this assumption and refer
to it as the \emph{isolation of exceptionals} hypothesis. 

Before considering ergodicity with respect to spin foams, it is useful
to first describe moves by which any admissible \emph{plaquette} configuration
can be reached from any other admissible plaquette configuration:

\begin{defn}[Plaquette cube move] 
A \emph{plaquette cube move} consists of
\begin{enumerate}
\item A choice of 3-cell of the cubic lattice. These are simply the
smallest cubes of the lattice, with 6 plaquettes, 12 edges, and 8
vertices.
\item For each of the 6 plaquettes of the 3-cell, a change is made to that
spin by $+\frac{1}{2}$ or $-\frac{1}{2}$ units of spin. 
\end{enumerate}
\end{defn}

It can be shown that these moves are ergodic on the space of admissible
plaquette configurations (ignoring for the moment non-trivial global
topology of the lattice%
\footnote{If boundary conditions result in a lattice of non-trivial homology,
homology changing moves must be added, as described below.%
}); a similar argument is used for instance in~\cite{BaezEtal}.

We next observe that, within our present context where the amplitude
depends on intertwiner labellings, the plaquette cube moves will transform
between \emph{admissible} plaquette configurations but \emph{inadmissible}
spin foams, unless some simultaneous changes are made to the intertwiners
labelling the edges of the cube. This leads us to define \emph{spin
foam} cube moves as follows: 

\begin{defn}[Spin foam cube move]
A \emph{spin foam cube move} consists of the
following:
\begin{enumerate}
\item A plaquette cube move.
\item For each edge in the 3-cell chosen for the plaquette cube move, a
randomly chosen change is made to the intertwiner labelling that edge, from a
fixed set of possible changes. The possible changes are such that for any 
plaquette cube move, there is always a change that results in an admissible spin foam. 
The intertwiner changes that are proposed depend on how the edges are split within the selected cube; further details can be found in Appendix~\ref{sse:compatible}.
\end{enumerate}
\end{defn}

We should mention here that for all the spin foam moves discussed
in this section there will be many moves that are rejected immediately
due to violation of the constraints; these are nonetheless counted as moves in
order to satisfy detailed balance. 

By construction, spin foam cube moves will reach some subset of intertwiner
labellings for each plaquette configuration. The possibility that
this may omit some admissible intertwiner configurations
leads us to introduce an independent edge move.

\begin{defn}[Spin foam edge move]
A \emph{spin foam edge move} consists of the
following:
\begin{enumerate}
\item An edge is selected.
\item The intertwiner label is incremented or decremented by two half-units
of spin (to preserve parity). 
\end{enumerate}
\end{defn}

To see that the combination of spin foam cube and edge moves are ergodic
on admissible spin foams (up to non-trivial global homology), we argue
as follows. Observe that the spin foam cube moves allow us to move
freely amongst the admissible plaquette configurations. To reach any
given spin foam, one first obtains the associated plaquette configuration
through the spin foam cube moves. We then apply whatever spin foam
edge moves are needed to set the edge labels to their given values.
This is possible because the admissible ranges $I_{e}$ are connected by 
spin foam edge moves. 

We now turn to the the issue of configurations arising from the non-trivial
global topology of the lattice. It is common in lattice computations
to impose periodic boundary conditions which are toroidal; i.e.\ in
the present case the lattice is a discretization of the $3$-torus.
For these boundary conditions, if one introduces a sheet
of half-unit spin having non-trivial global topology (see Figure~\ref{fig:visuals}, right), the spin foam cube moves
are not able to remove it. Such a sheet (and its deformations by cube
moves) can be introduced in any of the three directions of the 3-torus,
in correspondence with the three generators of the second homology group
of the 3-torus. A simple way to move between configurations corresponding
to different homology classes is to create and remove sheets of half-unit
spin; we do this by introducing the following moves: 

\begin{defn}[Spin foam homology move] 
A \emph{spin foam homology move} consists
of the following:
\begin{enumerate}
\item A 2-dimensional plane of plaquettes spanning the lattice is selected.
Applying periodic boundary conditions, this plane topologically defines
a 2-torus wrapped around the 3-torus. 
\item For the selected plane, change each edge and plaquette label contained in it 
by the same amount, either $+\frac{1}{2}$ or $-\frac{1}{2}$.
\end{enumerate}
\end{defn}

We note that the cost of computing the change in amplitude induced
by these moves scales with $L^{2}$ (where $L$ is the side length of the lattice), 
making them costly for large lattices; we shall return to this point in our discussion below. 

Now we have all the components to give a simple statement of our algorithm.

\begin{algorithm}
(Ergodic spin foam algorithm) \emph{A single iteration of the algorithm
consists of the following:}
\end{algorithm}
\begin{enumerate}
\item Apply a single spin foam cube move, spin foam edge move, or spin foam
homology move.
\item For the edges affected by the move, immediately reject the move if
any spin foam admissibility condition is violated.
\item Accept the move if  $\frac{\left|A_{new}\right|}{\left|A_{old}\right|}>\alpha$,
where $A_{new}$ and $A_{old}$ denote the spin foam amplitudes before
and after the move, and $\alpha$$ $ is a sample drawn
from from the uniform distribution on the unit interval.
\end{enumerate}
Step 1 gives a choice from among a set of moves that are ergodic on admissible
spin foams, not just within a homology class but between classes as well. For
each type of move, detailed balance is guaranteed as any move and
its inverse is proposed with equal probability. The relative probability
of proposing a cube, edge, or homology move can be freely chosen,
for example to tune the algorithm for better convergence or acceptance
rate. Step 2 could in principle be absorbed into Step 3, as the zeroes
of the $18j$ symbols enforce all the admissibility constraints. In
practice, admissibility is very easy to check and so one can often
avoid computing the full $18j$ symbol. Step 3 is just the usual Metropolis
acceptance condition; observables depending on the dual variables
can be measured at each iteration or at regular intervals. The absolute
value of the amplitudes in Step 3 are necessary due to the possibility
of negative amplitudes (arising from the $18j$ symbols). The standard
sign trick for computing expectation values in this situation is reviewed
in Section~\ref{ssse:sign-problem}.

In summary, we have defined a dual Metropolis algorithm for non-abelian
$SU(2)$ pure Yang-Mills in the case of three dimensions. Assuming
the isolation of exceptionals hypothesis, this algorithm is ergodic
on supported spin foams. Before presenting results for our algorithm
in Sections 3 and 4, we review some practical points of implementation.

\subsection{Algorithm Construction --- Practical considerations}

When using the above algorithm, there are a
number of novel considerations that don't apply to the conventional
case. We discuss some of the more important ones here.

\subsubsection{Spin foam versus plaquette configuration simulation}

It is possible to define an alternative dual algorithm based upon plaquette
configurations (rather than spin foams, which include intertwiner
labels), using the plaquette cube moves to move ergodically between
all admissible plaquette configurations. 
By applying the intertwiner expansion and fast evaluation of the $18j$
symbols, one can automate the computation of plaquette configuration
amplitudes. This would proceed by writing the partition function 
(\ref{eq:statesum}) in terms of an intertwiner independent amplitude: 
\begin{equation}
\mathcal{Z}=\sum_{j}\hat{A}(j),\quad\hat{A}(j)\equiv\left(\sum_{i}
\prod_{v\in V}{18j}^v(i_{v},j_{v})\prod_{e\in E} N^e(i_e,j_e)^{-1}\right)
\left(\prod_{p\in P}e^{-\frac{2}{\beta}j_{p}(j_{p}+1)}(2j_{p}+1)\right),
\label{eq:statesum2}
\end{equation}
where the amplitude $\hat{A}(j)$ is a function of plaquette labelling
$j$ only; i.e.\ the full sum over intertwiner labels has been absorbed
into the amplitude for a given plaquette labelling.

The question then becomes whether this algorithm is practical.
We found that the answer depends crucially on the strength
of the coupling. For comparison, we implemented the plaquette configuration
simulation and were able to achieve good agreement with conventional
results at \emph{sufficiently strong coupling} ($\beta<1.8$), where
configurations involving large intertwiner sums are penalized by the
amplitude. However, starting at $ $$\beta\approx1.8$ the simulation
began to spend time in plaquette configurations with a large space
of admissible intertwiner labellings; because the summations are nested
there are a huge number of edge labelled terms being evaluated for
a single plaquette configuration. Compounding this situation is a
second critical drawback, in that for many configurations (and increasingly
so at weaker coupling) the terms in the intertwiner summations couple
edges that are arbitrarily distant in the lattice. Thus, at large
lattice sizes and weak coupling the cost of computing the change in
amplitude due to a single move will on average be of the order of
the lattice volume.

The spin foam algorithm defined in Section 2.2 avoids both problems
by making local changes to both plaquettes and edge labels. In summary,
rather than importance sampling the plaquette configurations with
a generally expensive, non-local amplitude (the intertwiner sum),
one applies importance sampling to the innermost terms of~(\ref{eq:statesum}),
which carry both plaquette and edge labels. While this involves a larger
space of configurations, the evaluation of amplitudes proceeds extremely
rapidly and independent of coupling. The observation that such a local
algorithm exists for dual Yang-Mills theories (and also for dimensions
higher than 3) was made by Halliday and Suranyi in~\cite{Halliday95},
although they do not make explicit use of the spin foam formalism.

\subsubsection{Sign Problem}
\label{ssse:sign-problem}

Because the $18j$ symbol can take on negative values,
the most straightforward form of the Metropolis algorithm is not applicable.
Nonetheless, one can try to avoid this problem by applying what is
commonly referred to as the ``sign trick'', e.g.~\cite{deRaedt}.
We note that the sign trick was also employed in the work of Hari Dass~\cite{Dass83}.
Letting $\epsilon(f)$ denote the sign of the amplitude $A(f)$ 
for spin foam $f$, one can express the expectation value of an observable
$O(f)$ as follows:
\begin{equation}
\left\langle O\right\rangle =\frac{\sum_{f\in F^{+}}O(f)A(f)}
{\sum_{f\in F^{+}}A(f)}=
\frac{\sum_{f\in F^{+}}O(f)\epsilon(f)\left|A(f)\right|}{\sum_{f\in F^{+}}\epsilon(f)\left|A(f)\right|}
=\frac{\left(\frac{\sum_{f\in F^{+}}O(f)\epsilon(f)\left|A(f)\right|}
{\sum_{f\in F^{+}}\left|A(f)\right|}\right)}
{\left(\frac{\sum_{f\in F^{+}}\epsilon(f)\left|A(f)\right|}
{\sum_{f\in F^{+}}\left|A(f)\right|}\right)}=\frac{\left\langle
 \epsilon O\right\rangle _{||}}{\left\langle \epsilon\right\rangle _{||}}.
\label{eq:signtrick}
\end{equation}
Observe that we now have a ratio of two expectation values, which
can be estimated using simulations of a system governed by the \emph{absolute
value} $ $$\left|A(f)\right|$ of the original amplitude (we indicate
this by adding a subscript $||$). A common failing of this approach
is that for many systems studied in quantum Monte Carlo (where the
sign problem arises frequently), both numerator and denominator of
the expression decay exponentially with a parameter under study. In
such cases there are more sophisticated approaches available; invariably,
resolving a serious sign problem requires exploiting foreknowledge
of the distribution of positive and negative weights in the configuration
space. 

As discussed in Section 4, for the range of $\beta<2.85$ and lattice
sizes up to the maximum studied ($16^{3})$, the numerator and denominator
of~(\ref{eq:signtrick}) are of order unity. In this regime, the sign
problem is very tame and does not prevent accurate results from being
obtained.

\subsubsection{Cost of Homology Moves}

Due to the extended nature of the structures added and removed by
the homology moves (involving on the order of $L^{2}$ vertices and
plaquettes), the associated change in amplitude becomes increasingly
expensive to calculate for larger lattices. The situation for the
present case of $SU(2)$ Yang-Mills in three dimensions can be summarized
as follows. Homology changing moves are necessary to achieve agreement
with conventional code on very small lattices, such as the $2^{3}$
lattice discussed in Section~\ref{se:verification}. However, we have found in practice
that as the lattice size increases, the frequency with which homology
moves are attempted can be greatly decreased and ultimately set to
zero without affecting the result. This is because adding a sheet
of half-unit spin has a very low acceptance rate for larger lattices.
We should mention however that for different spin foam models, particularly
those which are in a deconfining phase, the non-trivial homology sectors
of the ensemble will likely make significant contributions for arbitrarily
large lattice sizes, and so cannot be neglected.

\section{Verification of Dual Metropolis Code on A Small Lattice}
\label{se:verification}

Before proceeding to simulations at larger lattice size, an implementation
of the dual algorithm described in Section 2 was verified on a $2^{3}$
lattice. For this lattice and all lattices discussed in the paper,
toroidal boundary conditions were applied. The observable measured
was $\frac{1}{N}\sum_{p\in P}j_{p}$, the average spin per plaquette. 
The expectation value for this observable is:  
\begin{equation}
\left\langle j\right\rangle =\frac{\sum_{f\in\mathcal{F}}\left(\frac{1}{N}
\sum_{p\in P}j_{p}\right)\mathcal{A}(f)}{\sum_{f\in\mathcal{F}}\mathcal{A}(f)}.
\label{eq:spinobs}
\end{equation}
From the translation
invariance of our lattice, this is a good estimator for the expectation
value of spin on a single plaquette. 

\begin{figure}[tbp]
\includegraphics{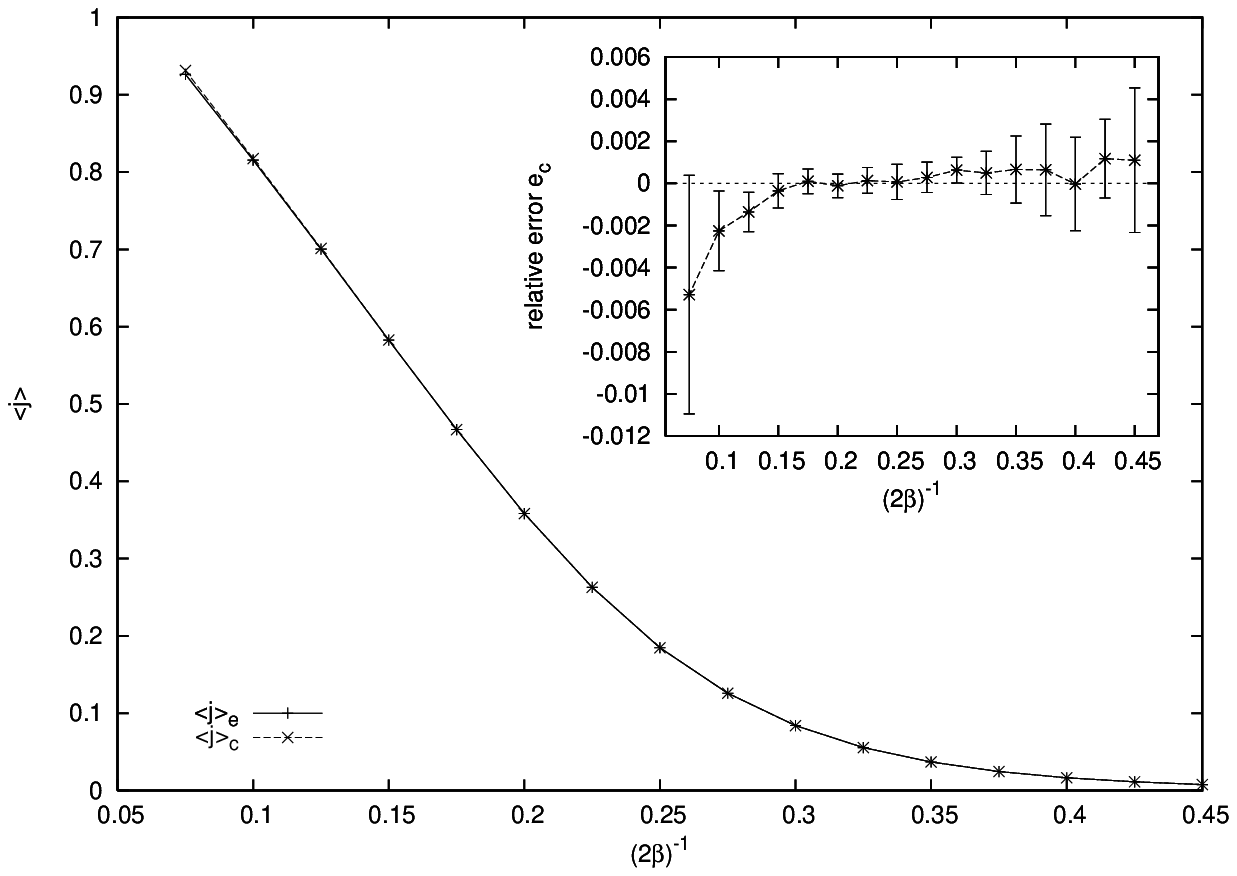}

\includegraphics{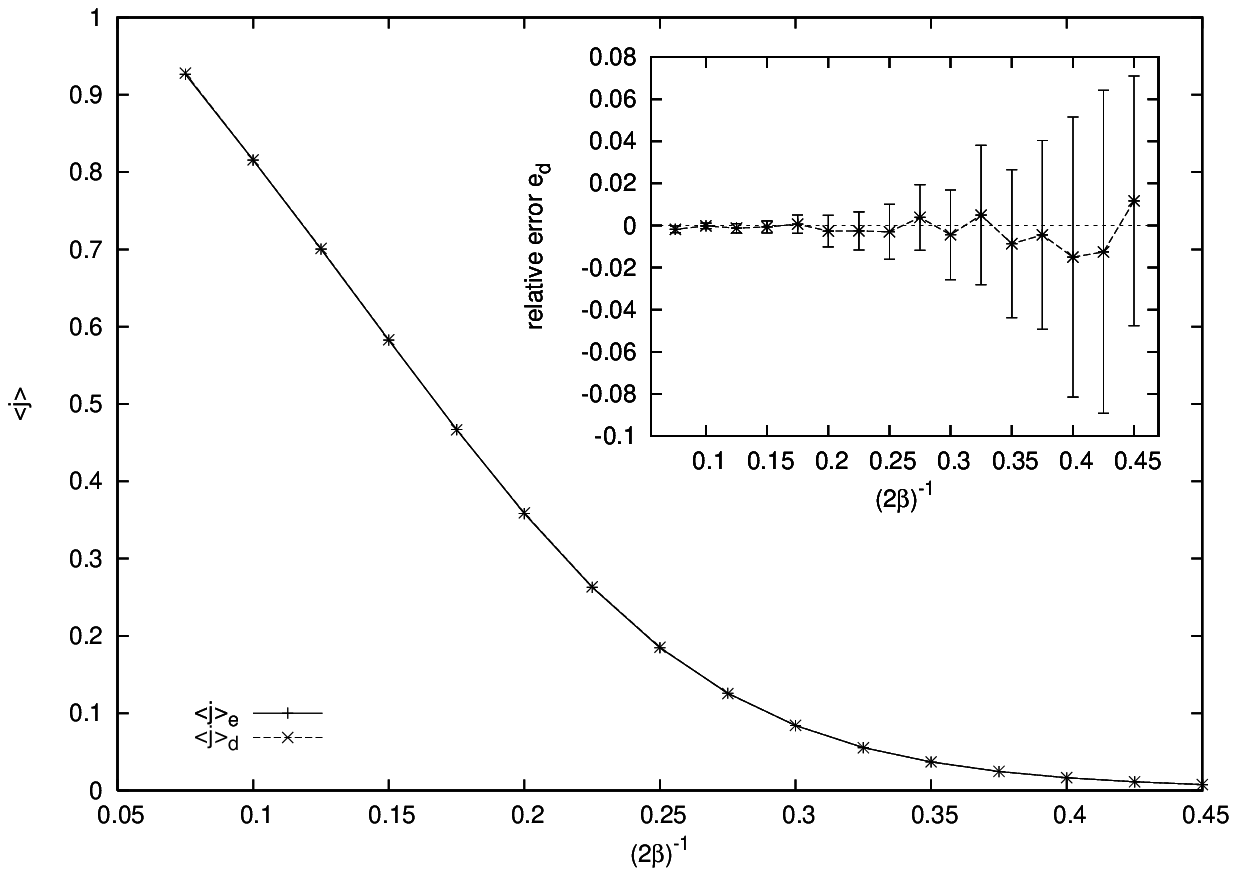}

\caption{Comparison of conventional (top) and dual (bottom) computations with
exact dual result on $2^{3}$ lattice for $j_{cut}=1$.}
\label{fig:exactcheck}
\end{figure}

While too small to be of physical interest, the $2^{3}$ lattice size
makes it easy to check agreement between different Monte Carlo algorithms,
as convergence is fast on smaller lattices. Moreover, by applying
a spin cutoff it is possible to \emph{exactly} compute the (cut-off)
partition function of the dual model on a $2^{3}$ lattice. This is
done for cutoffs $j_{cut}=\frac{1}{2}$ and $j_{cut}=1$, which correspond
to the restrictions that $j\in\left\{ 0,\frac{1}{2}\right\} $ and
$j\in\left\{ 0,\frac{1}{2},1\right\}$, respectively. While $j_{cut}=1$
gives on the order of $10^{10}$ configurations on the $2^{3}$ lattice,
the case of $j_{cut}=\frac{3}{2}$ already requires on the order of
$10^{15}$ configurations and is not practical. In Figure~\ref{fig:exactcheck}, 
we show
the results for spin cutoff $j_{cut}=1$.
For the conventional
computations,
the cut-off is realized by truncating 
the character expansion of the heat kernel; e.g.\ for $j_{cut}=1$
only the first three terms of the character expansion are used. 
This truncation leads to a mild sign problem for the conventional
calculation that can be resolved using the sign trick for the range
of $\beta$ presented here. 

The main plots in Figure~\ref{fig:exactcheck} each show two curves
which appear on top of each other.
The inset plots show the relative errors
\begin{equation}
\label{eq:relative-error}
  e_c \equiv \frac{\mean{j}_e - \mean{j}_c}{\mean{j}_e}
  \quad \text{ and } \quad
  e_d \equiv \frac{\mean{j}_e - \mean{j}_d}{\mean{j}_e},
\end{equation}
where the subscripts $c$, $d$ and $e$ refer to the conventional, dual
and exact calculations. The expectation values $\mean{j}_c$ and $\mean{j}_d$
and their standard deviations were found from the results of 12 
runs of $7.5\times10^{8}$ simulation steps each. The error bars are $3\sigma$ in magnitude.
The data is consistent with the expected error being zero.

\section{Testing The Dual Algorithm}
\label{se:testing}

\subsection{Results}

\begin{figure}[tbp]
\includegraphics{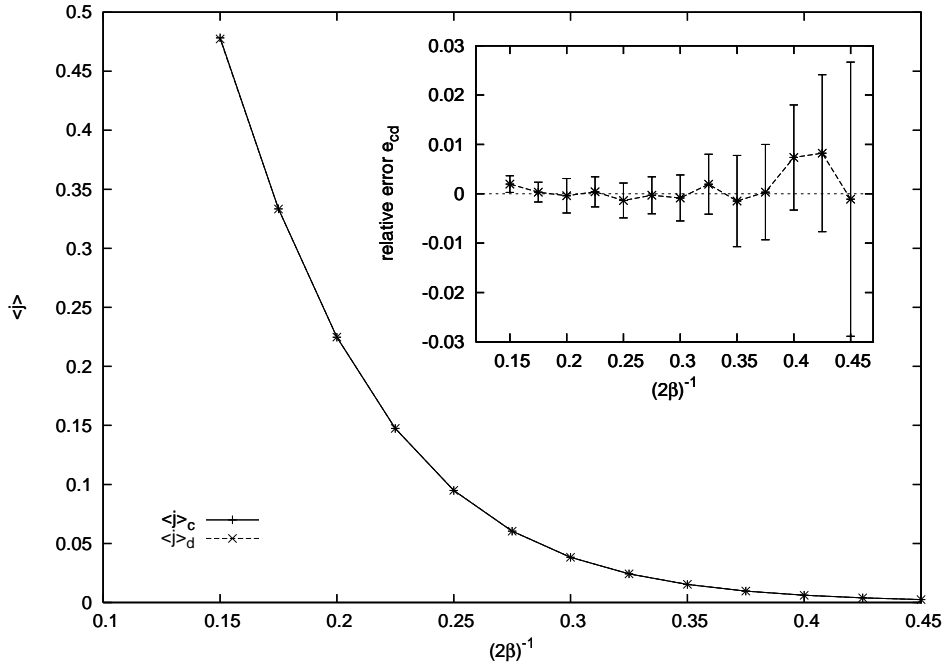}
\caption{Conventional and dual results for the $8^{3}$ lattice.}
\label{fig:L8_results}
\end{figure}

In this section we describe results obtained on an $8^3$ lattice 
over a range of coupling constants. For dual spin foam 
simulations, 25 runs of $10^{10}$ moves were used to generate the
data shown in Figure~\ref{fig:L8_results}; for the conventional results, 
15 runs of $10^{9}$ moves were used.
The observable measured was the
expectation value $\mean{j}$ of the spin observable, 
as described in Section~\ref{se:verification}. 
In the main part of Figure~\ref{fig:L8_results}, the expectation value 
for both conventional and dual results are shown, with standard deviation computed from
the results of different runs.  
Qualitatively, we see that the two algorithms are in very good agreement,
as the curves fall nearly on top of one another. To better exhibit
the detailed behavior, the inset plots show the 
relative error $e_{cd} \equiv \frac{\mean{j}_c - \mean{j}_d}{\mean{j}_c}$.
The standard deviation in the relative error was computed by
combining the standard deviations of the expectation values.
All error bars are $3\sigma$ in magnitude.
We find agreement between the two algorithms within about 1\%, 
which is within the estimated error.

The results for lattices of size $4^3$ and $16^3$ are very similar.
The above computations were done with the ladder recoupling of the
$18j$ symbol (see Appendix~\ref{ssse:ladder}) due to its simpler implementation.
For the range of coupling presented, we found that
the results were insensitive to increases in the cut-off past $j_{cut}$ = 3, which 
was the specific value used for both the dual and conventional data.  

\subsection{Slow-down at weak coupling}

With the current implementation of the algorithm, there is a problem
in obtaining results for $\beta\geq2.85$. For these $\beta$, the
Metropolis simulation eventually becomes ``trapped'' in a region
of higher spin than untrapped configuations.
These trapped regions appear to have steep or narrow exits,
as the simulation remains trapped for the duration of
the computation. This behavior is also
observed in some runs at stronger coupling (small $\beta$); the trapped mode generally
takes a longer time to develop as coupling is increased, and is absent
entirely from most of the runs at the strongest couplings considered. 

While the precise nature of this slow-down is not yet fully understood,
it shows some of the symptoms of similar problems noted for dual $U(1)$
in three dimensions. In~\cite{CoyleHalliday}, long autocorrelation
times (not seen with the conventional code at the same $\beta$) are
observed past a certain $\beta$. Slow-down of dual algorithms due
to overly local moves at weak coupling is also alluded to in~\cite{Zach96,Zach98,Zack98PhysRev}
for $U(1)$ in four dimensions.

By analogy with the existing literature on the dual $U(1)$ case,
it is possible that the solution will involve adding moves that simultaneously
change spins over regions that are larger than the minimal cube. A
useful analog to consider may be spin systems, where the variables
are also discrete and slow-down has been successfully addressed using
cluster algorithms~\cite{BinderLandau}. In our case, the presence
of admissibility and parity constraints tends to make it difficult
to find moves with reasonable acceptance rates. As is common with
cluster algorithms, the ``size'' of the extended moves will likely
need to be tuned with $\beta$.

\section{Outlook for further Development}

\subsection{Generalizations}

Although we have so far limited ourselves to $SU(2)$ pure Yang-Mills in
three dimensions, the algorithm we describe is in principle straightforward
to extend to more physically relevant models. 

\subsubsection{Wilson Loop Observables}

The theory behind Wilson loop observables in the dual spin foam picture
can be found in the work of~\cite{OecklPfeiffer}.
The basic challenge for computations is the fact that, in the dual, the Wilson loop
is a ratio of two partition functions with different admissibility conditions.
In a forthcoming paper, we describe how to extend the method described here to define a Metropolis algorithm for Wilson loop observables~\cite{CherringtonWilson}. 

\subsubsection{Four Dimensions}

In four dimensions on a hyper-cubic lattice, the $18j$ symbol is replaced
by its higher-dimensional analog%
\footnote{In the same way that the dual model~\cite{Anishetty91} and~\cite{Anishetty93}
is equivalent to a particular recoupling of the $18j$ symbol (see
Appendix~\ref{ssse:tet}), we expect the $32j$ to agree with past work on dual forms
of Yang-Mills in four dimensions~\cite{Halliday95}.%
}, a $32j$ symbol$ $ (involving 24 plaquettes and 8 edges); for the
heat kernel action the plaquette weights take the same form as in
(\ref{eq:statesum}). As the edges will have six incident plaquettes,
the triangle inequalities will be more involved. Assuming these differences
can be accounted for, we expect an algorithm with the same general
form will be feasible. 

Of particular interest in $D=4$ will be the behavior of autocorrelation
times and simulation slow-down as the critical point (not present
in three dimensions) is approached in the continuum limit of large
lattice and weak coupling.

\subsubsection{Higher $N$ Lie groups and their $q$-deformations}

The algorithm has direct generalizations to $SU(3)$ and even $SU(N)$.
Vertex amplitudes (generalizations of the $18j$ symbol) will likely
be more intricate, as for higher $N$ more than one discrete variable
is needed to label a given irreducible representation.
The basic structure of the algorithm in terms of moves on plaquette and intertwiner 
labels should go through, although admissibility constraints can be expected to take a more complicated
form. 

An interesting feature of the dual model is the ease with which $q$-deformed
gauge groups $SU_{q}(N)$ can be treated once the infrastructure for
computing with $SU(N)$ is in place. One illustration of this is recent
work~\cite{KhavkineQdeform} where computations are performed with
a $q$-deformed version of the Riemannian Barrett-Crane model of spin
foam quantum gravity. In four dimensions, where one expects to approach
a critical point in the continuum limit, it would be interesting to
consider how computations of correlation functions are affected by
$q$-deformation, for example whether or not distinct universality
classes emerge. 

\subsubsection{Other Extensions}

The above mentioned extensions of the method are straightforward.
There are also intriguing (and more involved) possibilities of extending
the method into the realm of dynamical fermions and to gravity. For
incorporating dynamical fermions into a dual model, we are investigating
an approach in which the polymer expansion of the fermion determinant
plays a key role. For computing pure Yang-Mills coupled to spin foam
quantum gravity, work by Oriti and Pfeiffer~\cite{OritiPfeiffer}
provides a useful starting point. Developments in the theory of dual
models~\cite{Oeckl,PfeifferDual} indicate that the dual viewpoint
has broad applications. We hope that concrete computational results
will encourage its further development and offer an alternative means
for analyzing theories of current interest. 

\section{Conclusions}

Previous work with computing in the dual of non-abelian theories due
to Hari Dass et al.~\cite{Dass83,Dass94,DassShin} brought to light a number
of important issues, but it had not been clear how these might be fully resolved.

The work reported here establishes that one can indeed obtain agreement
with conventional lattice gauge theory using simulation with the dual
form of the theory. Currently, the algorithm we describe has provided
results on a $16^{3}$ lattice for $\beta$ up to 2.85, and work is
underway to extend this to larger lattices and weaker coupling
(higher $\beta$). We expect addressing higher $\beta$ will involve
introducing moves that change spin foams over a larger region of the
lattice, as our algorithm appears to exhibit the symptoms of crossover
to a delocalized, disordered phase as observed in~\cite{CoyleHalliday,Zach96,Zach98,Zack98PhysRev}
for the case of dual three-dimensional $U(1)$. In four dimensions,
the expectation is that similar behavior will be seen as the critical
point is approached. The fact that cluster algorithms have been remarkably
successful at addressing critical slow-down in discrete spin systems
is promising in this regard and it is likely that some insights from
this area may apply.

The ultimate impact of the sign problem is still not fully understood
from the present work, as the sign expectation begins decreasing away
from unity at close to the same $\beta$ where the current simulation
is obstructed by impractically long autocorrelation times. Assuming
a practical resolution to the trapping problem can be found, it will
be interesting to see whether or not the sign problem becomes a serious
obstacle at weaker coupling.

An important property of our algorithm is the use of both plaquette
labels and edge (intertwiner) labels as variables of the simulation.
This allows the Metropolis algorithm to proceed through local moves
whose changes in amplitude can be evaluated very rapidly. Using the
spin foam formalism, ergodicity is fairly straightforward to show;
it is essentially an extension of an algorithm used in previous (quantum
gravity) spin foam computations~\cite{BaezEtal,KhavkineQdeform} to
the case where intertwiner labels are changed as well. Our claim of
ergodicity does however depend on the unproven hypothesis that exceptional
zeros of the $18j$ symbol are isolated, i.e.\ the zeros do not form
surfaces that separate the space of admissible spin foams. The agreement
of our results with conventional computations can be viewed as evidence
for this hypothesis.

Another interesting aspect of our algorithm is the necessity (for
full ergodicity) of introducing moves that change homology class,
in the case where boundary conditions lead to non-trivial global topology
of the lattice. While for the specific model studied here such configurations
have negligible weight for large lattices, in more general spin foam
models (e.g.\ those in a deconfining phase) these configurations may
play a more important role. 

Also worth emphasizing is the important role of diagrammatic techniques
in generating efficient algorithms for the $18j$ symbols. In particular,
the formula of Anishetty et al.~\cite{Anishetty91,Anishetty93} used
in~\cite{Dass83,Dass94,DassShin} can be derived very economically
from a particular splitting of the $18j$ symbol, as we show in the
Appendix. 

While still in its very early stages (particularly in comparison with
several decades development of conventional LGT codes), we believe
dual computations have a promising future ahead. The algorithm we describe here
immediately generalizes to higher $N$ gauge groups and to four dimensions,
and work is underway to incorporate dynamic fermions into the dual
framework. The next major step is the computation of Wilson loop observables,
currently in progress~\cite{CherringtonWilson}. Finally, the application
of the technique to spin-foam quantum gravity coupled to matter provides
a long-term motivation for the present work.

\begin{acknowledgement*}
The authors would like to thank Florian Conrady, N.D.~Hari Dass, Roman Koniuk, Robert Oeckl, 
and Hendryk Pfeiffer for valuable discussions. The first author was
supported by an NSERC postgraduate scholarship, the second by an
NSERC grant and the third by NSERC and FQRNT postgraduate scholarships. 
This work was made possible by the facilities of the 
Shared Hierarchical Academic Research Computing Network (SHARCNET).
\end{acknowledgement*}

\appendix
\section{The dual model, $18j$ symbol algorithms and compatible intertwiner moves}

\subsection{Derivation of the dual model}\label{sse:derivation}
This section sketches some of the steps of the transformation from the
conventional to the dual form of the lattice Yang-Mills partition
function, \eqref{eq:conventional} and \eqref{eq:statesum} respectively.
Our approach is inspired by the spin foam picture, and is closest 
to that found in \cite{Conrady}. Non-abelian dual models have also been analyzed
from a spin foam perspective in \cite{OecklPfeiffer,Oeckl}.

We begin by observing that, due to gauge invariance, the plaquette action $S(g_p)$ of
\eqref{eq:conventional} depends only on the conjugacy class of its
argument. Thus, its exponential can be expanded in terms of group characters $\chi_j$
\begin{equation}
	e^{-S(g)} = \sum_j c_j \chi_j(g),
\end{equation}
where $j$ ranges over the equivalence classes of finite-dimensional
irreducible unitary representations of the gauge group $G$.
Substituting into~\eqref{eq:conventional} and interchanging the order of summation and integration yields
\begin{equation}\label{eq:Zcharexp}
	\mathcal{Z} = \sum_{\{j_p\}} \int \prod_{e\in E} dg_e
		\prod_{p\in P} c_{j_p} \chi_{j_p}(g_p).
\end{equation}
At this point it is convenient to specialize to a $D=3$ cubic lattice
with periodic boundary conditions
and to fix an orientation for the plaquettes and edges of the lattice.
Choose a right-handed set of $xyz$ axes for the lattice.  Orient all
of the edges in the positive coordinate directions.
Every lattice cube is in the first octant of one of its vertices. Take each of
the three plaquettes of the cube that are incident to this vertex and
orient it in the counterclockwise direction, as seen from outside the
cube. It is easy to see
that this choice of orientations is translation invariant, that the
orientation of each edge agrees with two of the four plaquettes incident
on it and is opposite to the other two, and that every plaquette has
two edges whose orientations agree with its own and two that do not.

With this choice of orientation, the holonomy around a plaquette $p$ is
$g_p = g_1 g_2 g_3^{-1} g_4^{-1}$,
where $g_1, g_2, g_3$ and $g_4$ are the group elements associated to the
edges of the plaquette $p$, starting with an appropriate edge and
going cyclically.
Recall that the inverse $g_i^{-1}$ is used if the
orientation of edge $i$ does not agree with that of $p$.
Thus
\begin{equation}\label{eq:chi}
  \chi_{j_p}(g_p) = U_{j_p}(g_1)^b_a \, U_{j_p}(g_2)^c_b \,
                    U_{j_p}(g_3^{-1})^d_c \, U_{j_p}(g_4^{-1})^a_d \, ,
\end{equation}
where $U_j(g)^b_a$ denotes a matrix element with respect to a
basis of the $j$ representation.  If we insert~\eqref{eq:chi} 
into~\eqref{eq:Zcharexp} and collect together factors depending on
the group element $g_e$, we get a product of independent integrals
over the group, each of the form
\begin{equation}\label{eq:HIdef}
  \int dg_e \, U_{j_1}(g_e)^{b_1}_{a_1} \, U_{j_2}(g_e)^{b_2}_{a_2} \,
               U_{j_3}(g_e^{-1})^{b_3}_{a_3} \, U_{j_4}(g_e^{-1})^{b_4}_{a_4}
  = \int dg_e~\raisebox{-.70cm}{\includegraphics[height=1.8cm]{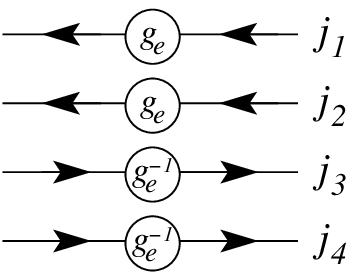}} \,\, .
\end{equation}
Here and below we use a graphical notation for tensor 
contractions, defined as follows.
Each \emph{wire} represents a matrix element of the unitary
representation labelling it.
Parallel wires represent products of such matrix elements.
The four matrix elements in~\eqref{eq:HIdef} come from the characters
associated to the four plaquettes incident on the edge $e$.
The free ends of the wires represent the indices of these matrix elements.
The wires can be joined together into loops, one for each plaquette.
The joining corresponds to contracting with other matrix elements 
from different edge integrals to form the product of characters as in~\eqref{eq:chi}.

Equation~\eqref{eq:HIdef} defines a projection operator on the
space of linear maps $j_{4} \otimes j_{3} \ra j_{1} \otimes j_{2}$.  
It is the usual group averaging operator whose image is precisely the 
intertwiners.
Since it is a projection operator, it can be resolved into a sum over a
basis of intertwiners $I_i: j_{4} \otimes j_{3} \ra j_{1} \otimes j_{2}$,
\begin{equation}\label{eq:HIresolution}
\int dg_e~\raisebox{-.75cm}{\includegraphics[height=1.8cm]{cables}} 
	~=~\sum_i \frac{I_i I^*_i}{\langle I^*_i,I_i\rangle}
	~=~\sum_i
		\frac{
			\includegraphics[height=1.5cm]{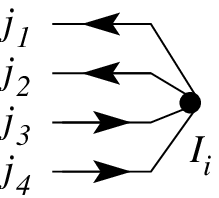} ~~
			\includegraphics[height=1.5cm]{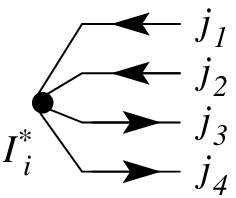}
		} {\raisebox{0cm}{\includegraphics[height=1.7cm]{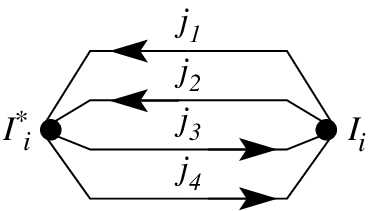}}},
\end{equation}
where the intertwiners 
$I^*_i:j_{1} \otimes j_{2} \ra j_{4} \otimes j_{3}$
are chosen such that the trace $\langle I^*_{i'},I_{i}\rangle$ of the 
composite $I^*_{i'} I_i$ is zero whenever $i'\ne i$ and non-zero if $i'=i$.
The projection property is readily verified.

If, for each edge of the lattice, we fix a term $i$ in the above
summation, we can contract the intertwiners $I_i$ and $I^*_{i}$
with those coming from the other edges.
At each vertex of the lattice, there will be six such intertwiners, and their
contraction can be graphically represented as an
octahedral network that we call the $18j$ symbol:
\begin{equation}\label{eq:oct}
\begin{matrix}\includegraphics[scale=0.7]{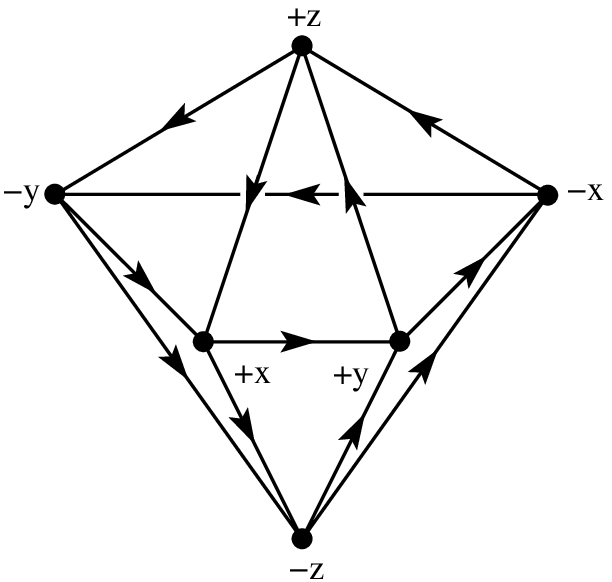}\end{matrix}.
\end{equation}
The vertices are labelled by the directions of the associated
lattice edges emanating from the given lattice vertex, namely 
$\pm x$, $\pm y$, and $\pm z$. 
The value of the $18j$ symbol depends on the choice of basis elements
$I_i$ and $I^{*}_{i'}$ in~\eqref{eq:HIresolution}, 
the six summation indices $i$ labelling the edges,
and the 12 incident plaquette labels $j$. 
Each normalization factor $N \equiv \langle I^*_i,I_i\rangle$
depends on the choice of basis elements at an edge, the
summation index $i$ on that edge, and the four plaquettes
incident on that edge.
Note that the choice of basis can be made independently
at each edge. 

The discussion up to this point has been quite general, assuming a
$3$-dimensional cubic lattice. Next, we specialize to $G=SU(2)$ and
give the plaquette action character expansion coefficients. For the
heat kernel~\eqref{eq:heatkernel}, the expansion coefficients take the
particularly simple form \cite{Menotti}
\begin{equation}
	e^{-S(g)} = \frac{1}{K(I,{\textstyle \frac{\gamma^2}{2}})}
		\sum_j (2j+1) e^{-\frac{\gamma^2}{2}j(j+1)} \chi_j(g),
		\quad {\textstyle j=0,\frac{1}{2},1,\ldots}~,
\end{equation}
where $K(g,t)$ is defined by~\eqref{eq:K}.
Putting these pieces together, we obtain the
dual formula for the lattice Yang-Mills partition function
\begin{equation}\label{eq:statesum-app}
	\mathcal{Z} = \sum_j \left(\sum_i
		\prod_{v\in V} 18j^v(j_v,i_v) \prod_{e\in E} N^e(i_e,j_e)^{-1}
		\right)\left(\prod_{p\in P} (2j_p+1) \, e^{-\frac{\gamma^2}{2}j_p(j_p+1)}\right),
\end{equation}
where an overall numerical factor of $K(I,\frac{\gamma^2}{2})$ 
per plaquette has been discarded.  
This precisely reproduces Equation~\eqref{eq:statesum}, where 
we described the notation we are using for the plaquette and edge
labellings $j$ and~$i$.

\subsection{Efficient algorithms for the $18j$ symbol via recoupling}
\label{sse:algs}

In order to perform computations with~\eqref{eq:statesum-app}, we
first need to choose explicit basis elements $I_{i}$ and $I^{*}_{i}$
of the spaces of intertwiners that appear in~\eqref{eq:HIresolution}.
Below, we consider two patterns for choosing such bases for each
edge of the lattice, one we call the ladder recoupling and one we call
the tetrahedral recoupling. They lead to different $18j$ symbols and have
different properties with respect to lattice translations.

\subsubsection{The ladder recoupling}\label{ssse:ladder}

Recall that for compatible spins $j$, $k$ and $m$, there is an
intertwiner $j \otimes k \ra m$ that is unique up to scale.  To be
explicit, we choose the specific intertwiner defined 
in~\cite[2.5.4]{CarterSaito}, and we denote it by
\begin{equation}
   \begin{matrix}\includegraphics{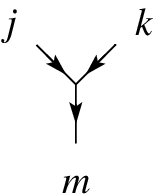}\end{matrix} \equiv
	 \begin{matrix}\includegraphics{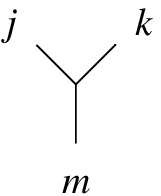}\end{matrix}.
\end{equation}
Similarly, we use the same reference\footnote{Note that our diagrams are read downwards, while those of~\cite{CarterSaito} are read upwards.}
 to define
\begin{equation}
   \begin{matrix}\includegraphics{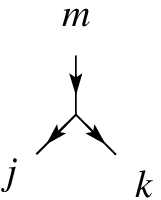}\end{matrix}
	 \equiv \begin{matrix}\includegraphics{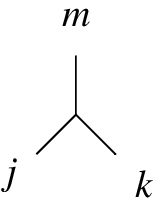}\end{matrix}.
\end{equation}
It is well-known that for fixed $j_{1}$, $j_{2}$, $j_{3}$ and $j_{4}$, the
intertwiners
\begin{equation}\label{eq:vert}
  I^{v}_{i} \equiv \begin{matrix}\includegraphics{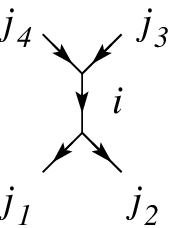}\end{matrix}
\end{equation}
form a basis of the space of intertwiners $j_{4} \otimes j_{3} \ra j_{1} \otimes j_{2}$,
as $i$ varies over admissible spins.
We call this the \emph{vertical splitting}.

There is also a second vertical splitting, given by interchanging
$j_{3}$ and $j_{4}$, which changes the intertwiner by a
factor of $(-1)^{j_{3}+j_{4}-i}$.
The geometry of the lattice provides a natural
way to choose between the two:  we make sure that the plaquette labels
on the left ($j_{4}$ and $j_{1}$ above) are part of the same lattice
cube, and same for the labels on the right.

A convenient choice of dual basis is given by
\begin{equation}
  I^{v*}_{i} \equiv \begin{matrix}\includegraphics{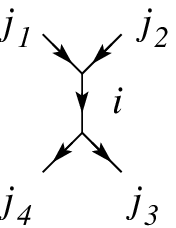}\end{matrix}.
\end{equation}
One can check that $\langle I^{v*}_{i'},I^{v}_{i}\rangle = 0$ for $i'\neq i$.
We next need to evaluate the normalization factor
\begin{equation}
  N^{v} = \langle I^{v*}_i, I^v_i\rangle = \raisebox{-.75cm}{\includegraphics[height=2cm]{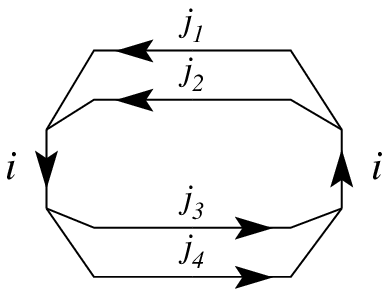}}.
\end{equation}
In order to accomplish this, we now explain how to relate our tensor
contraction diagrams to \emph{spin networks}.
While it would be possible to work entirely with tensor contraction
diagrams, there are two reasons to switch to the spin network notation.  
First, spin networks do not require that the edges be directed, which
relieves us of some complicated bookkeeping.
Second, by using spin networks, we can take advantage of many existing
formulas and software libraries for computing spin network
evaluations.

Recall that a spin network is a trivalent undirected ribbon graph
whose edges are labelled by spins.
One assigns a value to a spin network in the following way.
First, draw it in the plane, in general position, with the ribbon flat.
Then, read it from top to bottom, interpreting the vertices as the
trivalent intertwiners discussed above, and interpreting cups and
caps as certain intertwiners which can introduce signs.
If the spin network is closed, the resulting intertwiner is a map from 
the trivial representation to itself, and so can be identified with a
complex number.
The result is independent of the embedding in the plane, a fact that
is quite useful in computations.
In particular, the trivalent intertwiners are chosen carefully so
that the evaluation remains unchanged when inputs are deformed into 
outputs and vice versa.
We refer the reader to~\cite{CarterSaito} for more details.
While the starting point is different, the formulas given
in~\cite{KauffmanLins},
with $A=1$, also apply to these diagrams\footnote{Note that while 
we use half-integer spins, \cite{KauffmanLins} uses twice-spins, 
which are always integers.}.  
Note that some other authors have slightly different conventions, 
e.g.\ some take $A=-1$.

We will now work out how to compute tensor contractions using
spin networks.
Take a tensor contraction diagram involving just the trivalent
intertwiners discussed above and draw it in the plane such that
all edges are pointing downwards except for some edges which
leave the bottom of the diagram and loop around to reenter at
the top.  If we erase the arrows, the resulting spin network
will have the same interpretation as the tensor contraction
diagram, except for the signs introduced in the cups and caps.
One can show that the difference is exactly a factor of $(-1)^{2J}$,
where $J$ is the sum of the spins labelling the edges that loop
around.

As a first example, the value of a loop labelled with $j$ in
the tensor notation is the dimension $2j+1$ of the representation.
However, in the spin network notation, the value of a loop is
$\Delta_{j} = (-1)^{2j} (2j+1)$.

Similarly, the value of the edge normalization factor is
\begin{equation}\label{eq:Nvert}
  N^v = \langle I^{v*}_i, I^v_i\rangle 
  = \begin{matrix}\includegraphics[height=3.3cm]{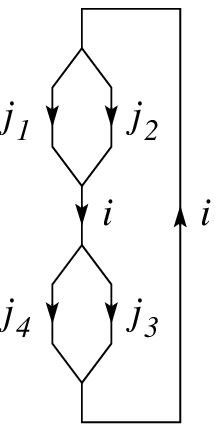}\end{matrix}
  = (-1)^{2i} \begin{matrix}\includegraphics[height=3.3cm]{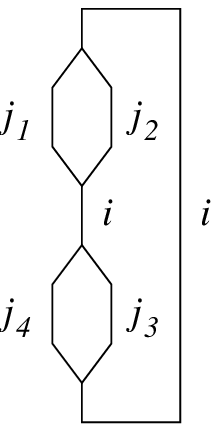}\end{matrix}
  = (-1)^{2i} \frac{\theta(j_1,j_2,i)\,\theta(j_3,j_4,i)}{\Delta_i},
\end{equation}
where $\theta(a,b,c)$ stands for the value of the
following \emph{theta network}:
\begin{equation}
	\theta(a,b,c) \equiv \begin{matrix} \includegraphics[scale=0.85]{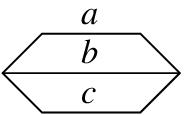} \end{matrix}.
\end{equation}
Its value is given explicitly in~\cite[Chapter 9]{KauffmanLins}.

Note that the conversion sign factor $(-1)^{2i}$ from~\eqref{eq:Nvert}
can be expressed as $(-1)^{2(j_1+j_2)}$ by appealing to the parity
constraints.
Since each plaquette is ``outgoing'' from two edges, each plaquette
spin contributes $(-1)^{4j} = 1$.  In other words, 
the conversion sign factors from the edge normalizations $N^v$ cancel.
It can be shown that the conversion sign factor for the
$18j$ symbol appearing in~\eqref{eq:oct} is independent of the edge
splitting and can be written as $(-1)^{2J}$, where, for instance,
$J=j_{{+x}{+y}}+j_{{+x}{-z}}+j_{{-y}{-z}}$. 
Each plaquette spin shows up in exactly one such sign factor, 
so the signs combine to give $(-1)^{2J_\text{tot}}$, where
$J_\text{tot}$ is the sum of all plaquette spins.  Note that
on a lattice with two or more odd side-lengths, this sign factor
can be non-trivial.

\begin{figure}[htbp]
\includegraphics[scale=0.625]{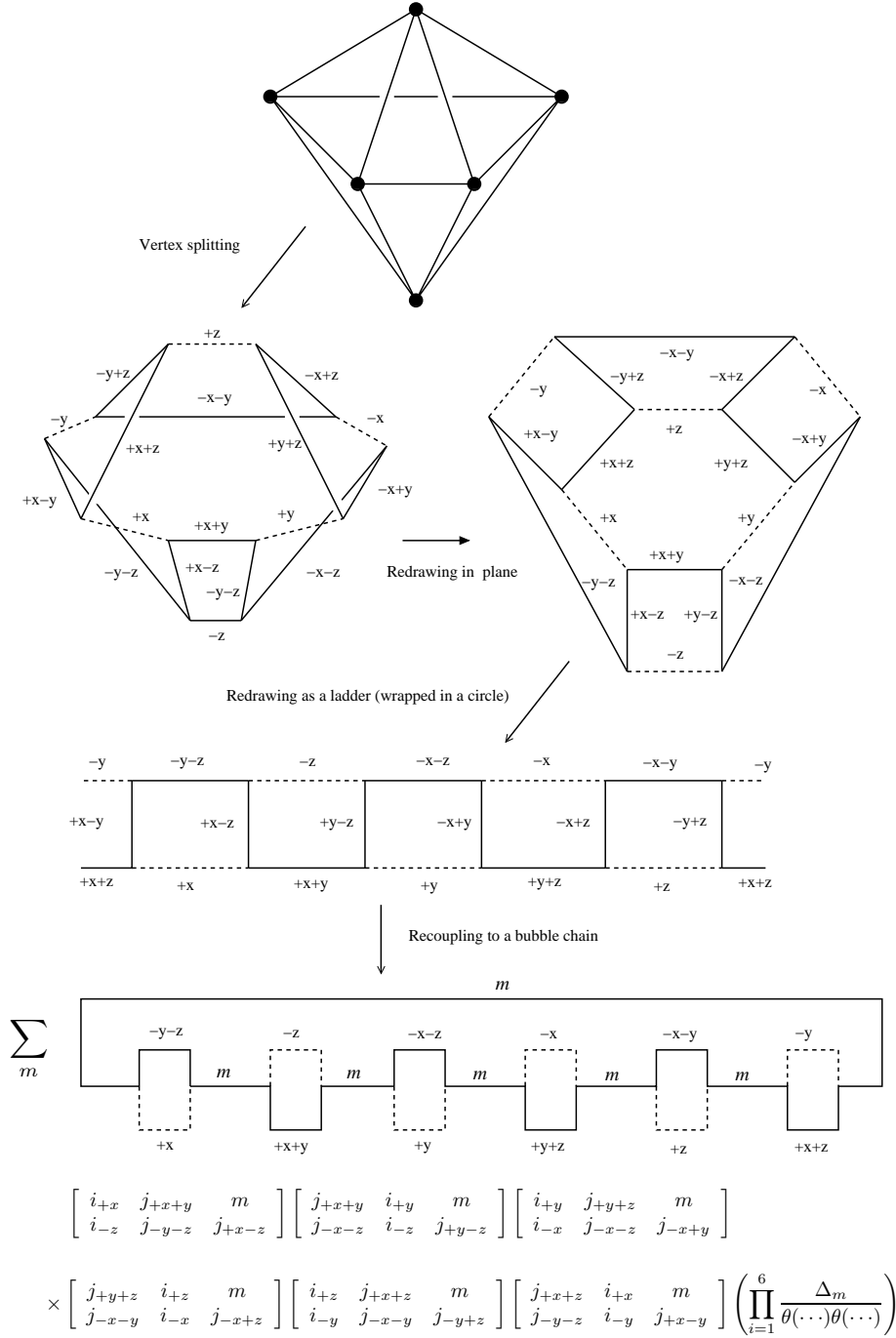}
\caption{The ladder recoupling of the $18j$ symbol; a single sum. }
\label{fig:ladrecoupling}
\end{figure}

Next we must work out the value of the $18j$ symbols that arise using
the vertical splitting. The corresponding spin network is obtained by
applying this splitting to the vertices of the octahedron shown
in~\eqref{eq:oct} and erasing the arrows from its edges.
A method for evaluating this spin network is shown in
Figure~\ref{fig:ladrecoupling}. 
The calculation is similar to
that of~\cite{ChristensenEgan}, where a ``ladder'' structure also appears. 
The recoupling move
\begin{equation}\label{eq:recouple}
\parbox{5in}{ \centerline{\includegraphics[scale=0.40]{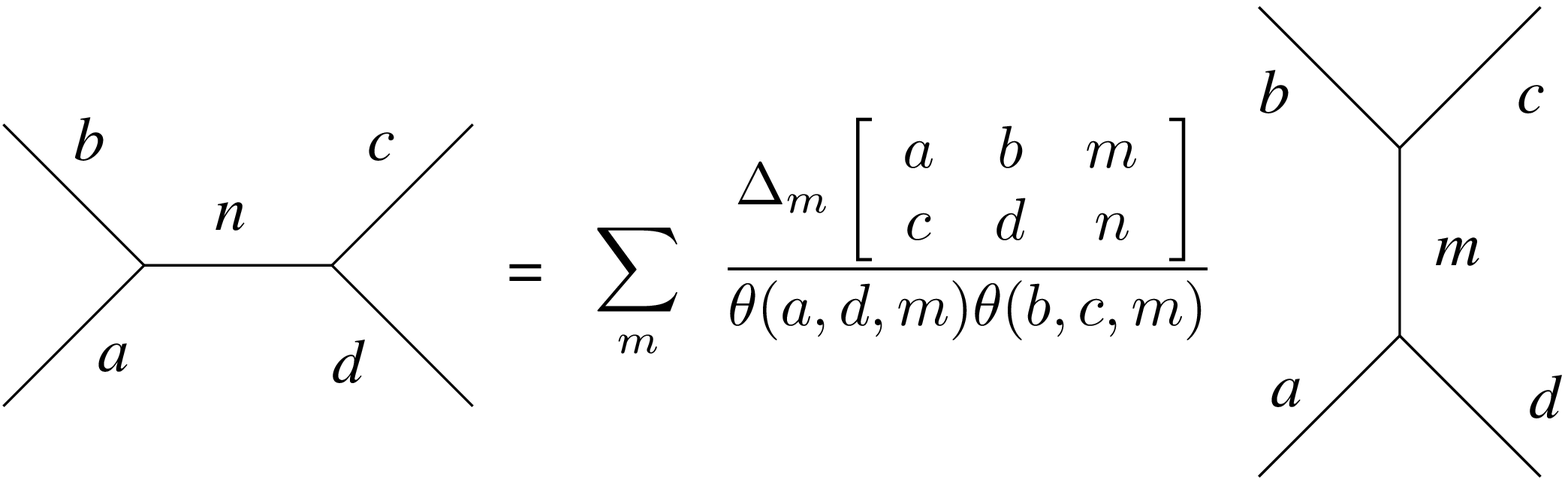}} } 
\end{equation} 
is applied to each of the six ``rungs'' of the ladder, 
producing a chain of bubbles.
The function of six spin labels appearing in~(\ref{eq:recouple}) 
is the tetrahedral network, shown in the last step of
Figure~\ref{fig:tetrecoupling}. The value of the tetrahedral network is
given explicitly in~\cite[Chapter 9]{KauffmanLins} and is closely
related to the Wigner-Racah $6j$ symbol of angular momentum theory~\cite[Appendix B]{Messiah},
see~\eqref{eq:tet-6j}.

Because of Schur's Lemma,  the six independent sums from the
recoupling moves become a single sum.
The bubbles are proportional to the identity, weighted by a theta
network divided by a loop. Six theta networks arising from the bubbles cancel
against six of the twelve theta networks from the recoupling moves to give the six theta networks
shown in the final line.  The bubbles also contribute six loop factors ($\Delta_i$) in the denominator, which exactly cancel the loop factors from the recoupling.
The final result can be written as:
\begin{equation}
\sum_{m}\frac{\displaystyle
\raisebox{-2pt}{$\Delta_{m}$}\, \prod_{\text{6 bubbles}}\begin{bmatrix}
\cdots & \cdots &   m\\
\cdots & \cdots & \cdots\end{bmatrix}
}
{\mbox{\small$
\theta(m,i_{+x},j_{-y-z})\,\theta(m,i_{-x},j_{+y+z})\,\theta(m,i_{+y},j_{-x-z})\,\theta(m,i_{-y},j_{+x+z})\,\theta(m,i_{+z},j_{-x-y})\,\theta(m,i_{-z},j_{+x+y})
$}},
\end{equation}
where the arguments of the six tetrahedral networks are those that appear in the last line of Figure~\ref{fig:ladrecoupling}.
The explicit relation between the tetrahedral
network and the Wigner-Racah $6j$ symbols is
\begin{equation}\label{eq:tet-6j}
\begin{bmatrix}{J_1}&{J_2}&{J_3}\\{j_1}&{j_2}&{j_3}\end{bmatrix} =
\sqrt{|\theta(J_1,J_2,j_3)\theta(j_1,j_2,j_3)
       \theta(J_1,j_2,J_3)\theta(j_1,J_2,J_3)|}
\begin{Bmatrix}{j_1}&{j_2}&{j_3}\\{J_1}&{J_2}&{J_3}\end{Bmatrix} 
.
\end{equation}
Note the row swap and the fact that the four theta networks
correspond to the four triples of spins from the $6j$'s arguments that
must satisfy triangle inequalities. For reference, $|\theta(a,b,c)| =
(-1)^{a+b+c}\theta(a,b,c)$.

The $18j$ symbol described in this section was
used in computing the data appearing in Sections~\ref{se:verification}
and~\ref{se:testing}.

\subsubsection{The tetrahedral recoupling}\label{ssse:tet}

Next we consider a different splitting of the vertices of the
octahedron, which we call the tetrahedral recoupling.  The $18j$
symbol that arises here is more efficient to compute than the $18j$
symbol for the ladder recoupling, because it does not require a sum.
However, the splitting is not translation invariant, which makes it
slightly harder to work with.  This section is not needed in the rest
of the paper, but is useful as a comparison to other sources and
will be important for future calculations.

We begin by considering a different basis for the space of
intertwiners $j_{4} \otimes j_{3} \ra j_{1} \otimes j_{2}$.
It is given by the \emph{horizontal splitting}
\begin{equation}\label{eq:hor}
  I^{h}_{i} \equiv \begin{matrix}\includegraphics{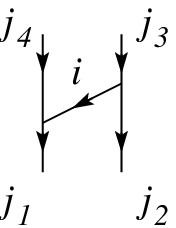}\end{matrix}
\end{equation}
as $i$ varies over admissible spins.
Note that it makes no difference which way the arrow on the edge
labelled by $i$ points.

There is also a second horizontal splitting, given by interchanging
$j_{3}$ and $j_{4}$.  As we did for the vertical splitting, we choose
between the two by requiring that $j_{4}$ and $j_{1}$ label plaquettes
that are part of the same cube.  Unlike the vertical splitting, 
the two horizontal splittings are not in general related by a sign.

A convenient dual basis is given by
\begin{equation}
  I^{h*}_{i} \equiv \begin{matrix}\includegraphics{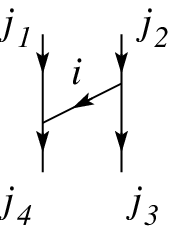}\end{matrix}.
\end{equation}
The traces $\langle I^{h*}_{i'},I^{h}_{i}\rangle$ work out to be
\begin{equation}\label{eq:Nhoriz}
	N^h = \langle I^{h*}_{i'},I^{h}_{i}\rangle
		= (-1)^{2(j_1+j_2)}\frac{\theta(j_1,j_4,i)\,\theta(j_2,j_3,i)}{\Delta_i}.
\end{equation}

\begin{figure}[tbp]
\includegraphics[scale=0.63]{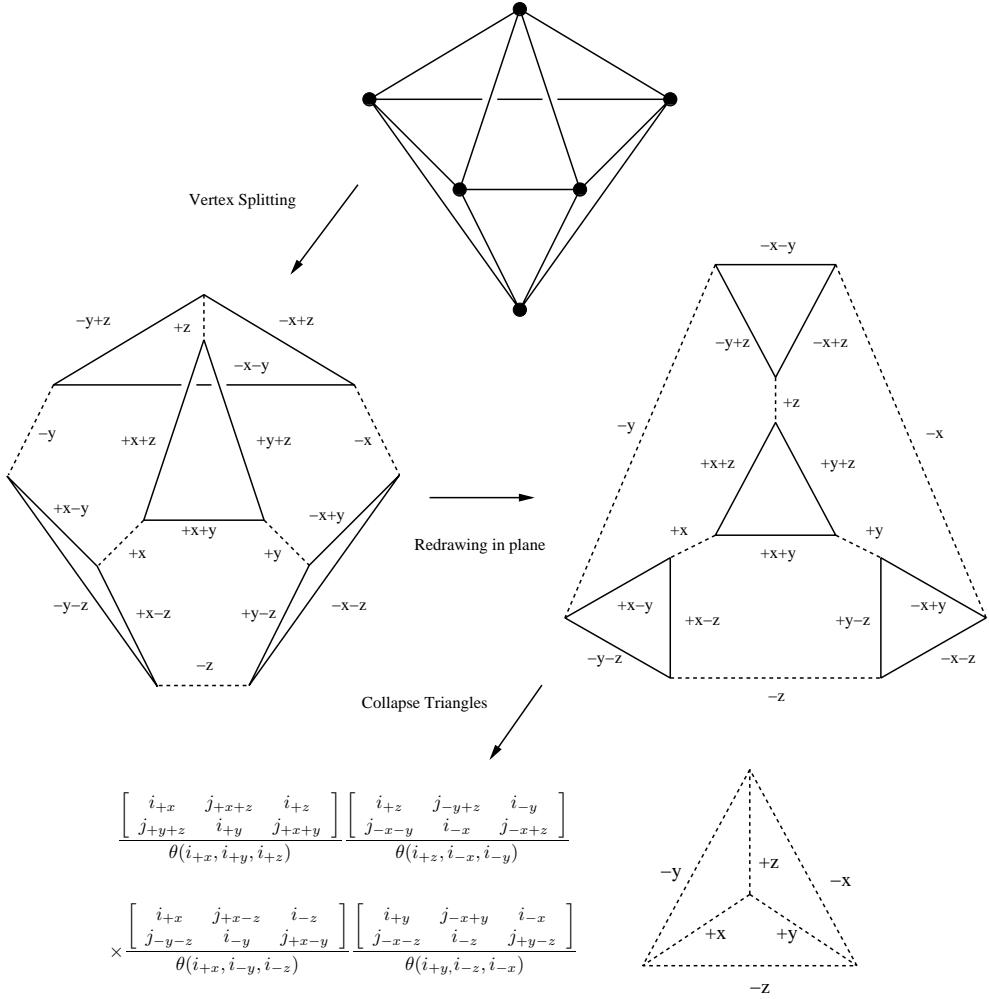}
\caption{The tetrahedral recoupling of the $18j$ symbol; no sum. }
\label{fig:tetrecoupling}
\end{figure}

For the tetrahedral recoupling, we use the vertical
splitting~\eqref{eq:vert} on three edges and the horizontal splitting
on the opposite edges.
The pattern we use is indicated in the first step of
Figure~\ref{fig:tetrecoupling}.
The result is a planar network consisting of four triangles connected to
one another in a tetrahedral pattern. 

For this recoupling, a minor complication
arises because the intertwiner splittings are different on opposite
edges. This means that a simple translation of the given tetrahedral
$18j$ symbol to neighboring vertices does not correspond to a consistent
choice of basis for the intertwiners. This is easily overcome
by dividing the lattice into a checkerboard of odd and even sites,
and alternately using the original and reflected versions of the $18j$
symbol. For periodic boundary conditions, this does limit one to 
lattices with even side-lengths, but this constraint is not serious in practice.

As was the case with the ladder recoupling, the conversion sign factors 
from the normalization factors cancel, and the conversion sign factors from the $18j$ symbols
give a factor of $(-1)^{2J_\text{tot}}$.  In this case, because the
checkerboard pattern forces even side-lengths, one can show that
$(-1)^{2J_\text{tot}} = 1$.

The diagrammatic relation
\begin{equation}
\parbox{5in}{ \centerline{\includegraphics[scale=0.64]{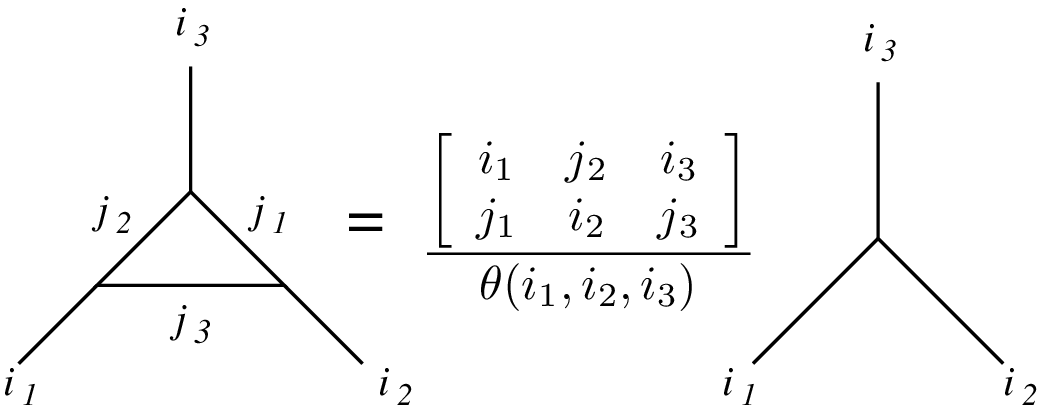}} } 
\label{eq:collapse}\end{equation}
can be used to collapse each of the four triangles into 3-valent vertices,
as shown in the last step of Figure~\ref{fig:tetrecoupling}.
The final result is simply:\smallskip
\begin{equation}\label{eq:tet-final}
\frac{\small
\begin{bmatrix}
i_{+x} \!\!\!& j_{+x+z} \!\!\!\!& i_{+z}\\
j_{+y+z} \!\!\!\!& i_{+y} \!\!\!& j_{+x+y}\end{bmatrix}
\!\begin{bmatrix}
i_{+z} \!\!\!& j_{-y+z} \!\!\!\!& i_{-y}\\
j_{-x-y} \!\!\!\!& i_{-x} \!\!\!& j_{-x+z}\end{bmatrix}
\!\begin{bmatrix}
i_{+x} \!\!\!& j_{+x-z} \!\!\!\!& i_{-z}\\
j_{-y-z} \!\!\!\!& i_{-y} \!\!\!& j_{+x-y}\end{bmatrix}
\!\begin{bmatrix}
i_{+y} \!\!\!& j_{-x+y} \!\!\!\!& i_{-x}\\
j_{-x-z} \!\!\!\!& i_{-z} \!\!\!& j_{+y-z}\end{bmatrix}
\!\begin{bmatrix}
i_{-x} \!\!\!& i_{-y} \!\!\!& i_{-z}\\
i_{+x} \!\!\!& i_{+y} \!\!\!& i_{+z}\end{bmatrix}
}
{
\theta(i_{+x},i_{+y},i_{+z})\,\theta(i_{+z},i_{-x},i_{-y})\,\theta(i_{+x},i_{-y},i_{-z})\,\theta(i_{+y},i_{-z},i_{-x})
}
\smallskip
\end{equation}
Because this formula is essentially a product of tetrahedral networks with no auxiliary summations, it is
highly efficient to compute. 

The tetrahedral recoupling is easy to express in terms of Wigner-Racah
$6j$ symbols using relation~\eqref{eq:tet-6j}.
In order to compare our work to other work, we give some of the details.
Ignoring signs for the moment, the theta networks from the 
edge normalizations~\eqref{eq:Nvert} and~\eqref{eq:Nhoriz}, 
the conversion formula~\eqref{eq:tet-6j}, 
and the vertex amplitude~\eqref{eq:tet-final} all cancel.
Now we collect the signs of the theta networks.
The theta networks from the edge normalizations contribute a sign of 
$(-1)^{2i} (-1)^{j_1+j_2+j_3+j_4}$, where $i$ labels the edge and the $j_k$ 
label the incident plaquettes. Since each plaquette is shared by four edges, 
the factors $(-1)^{j_1+j_2+j_3+j_4}$ cancel.
Thus the edge normalizations become $(-1)^{2i}/\Delta_i = 1/(2i+1)$.
Since this is positive, we can multiply each vertex amplitude by $\sqrt{2i+1}$
to take this into account.
The theta networks from~\eqref{eq:tet-final} contribute a sign of $(-1)^{\sum_{k=1}^6 2i_k}$,
where the $i_k$ label the edges incident on the vertex. Since each edge is
shared by two vertices, the vertex signs also cancel.
The final answer is that the vertex amplitude~\eqref{eq:tet-final} becomes
a product of five Wigner-Racah $6j$ symbols multiplied by a product of six
factors of the form $\sqrt{2i+1}$.

We observe that in this form the tetrahedral recoupling is equivalent to the dual
amplitude formula first proposed by Anishetty et al.~\cite{Anishetty91,Anishetty93}
and later used by Diakonov and Petrov~\cite{DiaPet99}. 
The same formula was used in the computational work of Dass~\cite{Dass83,Dass94,DassShin}.
It should be emphasized that previous derivations of this formula
did not make use of the spin foam formalism. As such, the identification
of extra labels (those not coming from original plaquettes) with intertwiners
was not explicit. We found this distinction between plaquette and intertwiner 
labels to be a crucial one in constructing our algorithm.

\subsection{Compatible intertwiner moves}
\label{sse:compatible}

As mentioned in Definition 2.4, when a (single) spin foam cube move
is applied, the 12 intertwiner labels of the selected cube have to
be adjusted to ensure the result is an admissible spin foam. In general,
admissible intertwiner changes depend on plaquette and edge labels of
the cube. However, we will show next that this dependence takes on a simple form
determined by the edge splittings. 

\begin{figure}[tbp]
\includegraphics[scale=0.55]{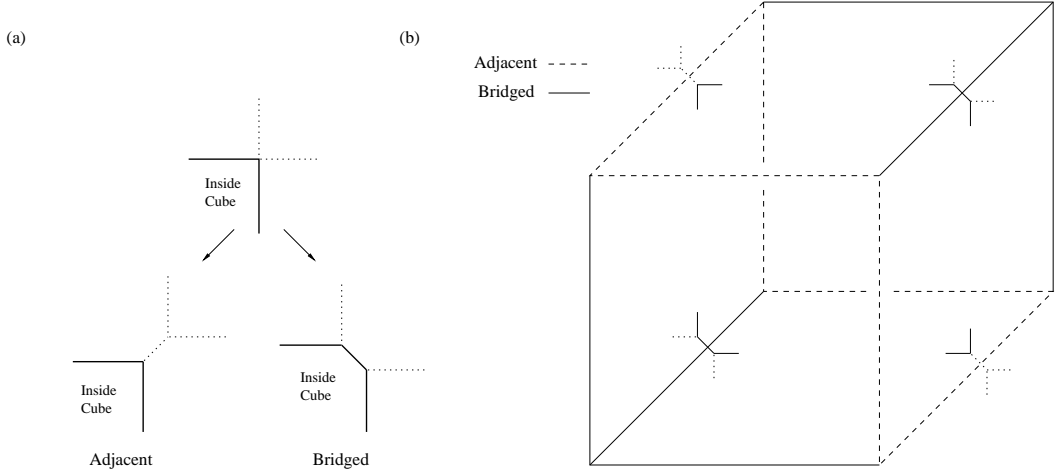}
\caption{ (a) The two types of splitting for an edge in a cube (side view).
(b) Two possible splittings associated with the ladder recoupling for the edges of a cube.}
\label{fig:itwiners}
\end{figure}

For this discussion, we consider the case where the edges are split according
to the ladder recoupling of the $18j$ symbol (see
Figure~\ref{fig:ladrecoupling}), as the translation 
invariance of this recoupling simplifies the analysis. 
From the point of view of a cube move, 
edge splittings come in two kinds, see Figure~\ref{fig:itwiners}(a). 
In one case, octahedral edges corresponding to adjacent
cube plaquettes remain \emph{adjacent} after the intertwiner edge is
introduced (type A). In the other, octahedral edges corresponding to
adjacent cube plaquettes become \emph{bridged} by the new intertwiner
edge (type B). The pattern of type A and type B edges for a cube
is shown in Figure~\ref{fig:itwiners}(b).

The main constraint on the admissibility of intertwiner changes is parity,
cf.\ Definition~2.2. For type A splittings, the two adjacent plaquette
spins change by $\pm\frac{1}{2}$, forcing the corresponding intertwiner
label to change by an integer. For type B splittings, the intertwiner,
now bridging the two adjacent plaquettes, must change by a
half-integer instead. A minimal set of admissible intertwiner changes
now consists of $1$, $0$ or $-1$, for a type A edge, and $\frac{1}{2}$
or $-\frac{1}{2}$, for a type B edge. Note that negative and positive
changes should be proposed with equal probability, to satisfy detailed
balance.

In summary a \emph{spin foam cube move} consists of the following:
\begin{enumerate}
\item Choose a cube in the lattice.
\item For each plaquette in the cube, adjust the spin independently
by $+\frac{1}{2}$ or $-\frac{1}{2}$.
\item For each type A edge in the cube, adjust the spin independently
by $1$, $0$ or $-1$.
\item For each type B edge in the cube, adjust the spin independently
by $+\frac{1}{2}$ or $-\frac{1}{2}$.
\end{enumerate}

%

\end{document}